\documentclass[aps,prd,twocolumn,showpacs,nofootinbib,amsmath,amssymb,amsfonts,showkeys]{revtex4-1}

\usepackage{tikz}
\usepackage{graphicx,import}
\usepackage{color, colortbl}
\usepackage{animate}
\usepackage{longtable}
\usepackage{lipsum}
\definecolor{LightCyan}{rgb}{0.88,1,1}

\begin{document}
\title{Thermal and Resonant Emission \\ of Dark Ages Halos in the Rotational Lines of HeH$^{+}$}

\author{Yu. Kulinich$^{1}$, B. Novosyadlyj$^{2,1}$, V. Shulga$^{2,3}$, W. Han$^{2}$}

\affiliation{$^{1}$Astronomical Observatory of Ivan Franko National University of Lviv, Kyryla i Methodia str., 8, Lviv, 79005, Ukraine;}
\affiliation{$^{2}$College of Physics and International Center of Future Science of Jilin University, Qianjin Street 2699, Changchun, 130012, People's Republic of China;}
\affiliation{$^{3}$Institute of Radio Astronomy of NASU, 4 Mystetstv str., 61002 Kharkiv, Ukraine 0000-0001-6529-5610}
%\collaboration{(LaTeX collaboration)}

\date{\today}
\begin{abstract}
 
We analyse the thermal emission and resonant scattering of CMB radiation from the dark ages halos in the rotational lines of the helium hydride ion (HeH$^{+}$), one of the first chemical compounds in the Universe. Evaluating the optical depth, thermal and resonant brightness temperatures and spectral fluxes of dark ages halos is based on computing the cross-sections and rate coefficients of excitation/de-excitation of the lowest five rotational states of HeH$^{+}$ by inelastic collisions with atomic hydrogen. It was shown that in the Dark Ages the collisional excitation/de-excitation by atoms of neutral hydrogen and electrons are competitive, whereas in the denser regions, e.g., in the halos, the contribution of collisions with atomic hydrogen is dominant. We demonstrate the peak-like time-dependence of halo luminosities in HeH$^{+}$ lines.

\end{abstract}
%\pacs{95.36.+x, 98.80.-k}
\maketitle

\section{Introduction}

In astrophysics, the interest in the helium hydride ion HeH$^{+}$ discovered in the lab in 1925 \cite{Hogness1925}, is caused by its stability in the cosmic isolation and by its composition, since it consists of the most widespread atoms in the Universe - hydrogen and helium. It is not surprising that numerous discussions about the formation of this molecule and its search both in the Galaxy and the deep space have been undertaken since the 70s of the last century \cite{Jura1971,Dabrowski1978,Frommhold1978,Roberge1982,Zygelman1998,Kimura1993,Jurek1995,Liu1997, Kraemer1995,Zygelman1998,Engel2005,Zinchenko2011}. Unfortunately, the HeH$^{+}$ detection has not directly been confirmed for a long time and only recently the first reliable confirmation of the HeH$^{+}$ existence towards the planetary nebula NGC 7027 has been obtained \cite{Gusten2019}.

Molecule HeH$^{+}$ plays a significant role in the formation of the first stars at the end of Dark Ages since it is one of the first molecules to appear in the early Universe among with H$_2$, HD, LiH, and their ions \cite{Dubrovich1977,Puy1993,Puy1996,Puy2007,Dubrovich1997,Galli1998,Lepp2002,Dalgarno2005,Hirata2006,Novosyadlyj2018}. Molecules are coolers of baryonic matter in the early Universe at temperatures below $\sim$8000 K \cite{Galli1998}. Emissions of these molecules result in cooling, fragmentation, and collapse of molecular clouds. In particular, molecular ions HeH$^{+}$ is an effective cooler \cite{Coppola2011} due to the large value of their electric dipole moment, $1.722$~D, emitting in the molecular lines in the ground vibrational and electronic state. The population of the rotational levels in the epoch of Dark Ages is defined by collisions with photons of the cosmic microwave background (CMB), free electrons, ions and atoms.
Calculations of coefficients for rotational and vibrational excitation/de-excitation rates by electrons are performed by \cite{Rabadan1998,Hamilton2016,Curik2017,Khamesian2018}.
However, up to now, there are no estimations for HeH$^{+}$ rotational excitation/de-excitation by collisions with atomic hydrogen.

The rest of the paper is organized as follows: In Section II, the potential energy surface (PES) for H--HeH$^{+}$ collisions is obtained in analytical and numerical approximations. In Section III, the state-to-state integral cross sections for rotational transitions during H--HeH$^{+}$ collisions are obtained and corresponding rate coefficients are calculated. 
The results for rate coefficients are presented in graphical form and analytical approximations. In Section IV, the role of collisional excitations of low rotational levels of HeH$^{+}$ in Dark Ages is discussed. In Section V, the differential thermal and resonant brightness temperatures, as well as spectral fluxes from dark ages halos in rotational lines of the helium hydride ion HeH$^{+}$ are evaluated. The conclusions are in Section VI.

\section{PES for H-$\textrm{HeH}^{+}$ interaction} \label{sec:style}

\begin{figure}
\begin{tikzpicture}[domain=0:2]
\draw[->,line width=0.5mm,dashed] (-1,0.5) -- (5.5,0.5) node[below right] {$r_{xy}$};
\draw[->,line width=0.5mm,dashed] (0,-1) -- (0,5.5) node[left] {$z$};
\draw[->,line width=0.5mm] (0,0.5) -- (4.5,4.5);
\draw[line width=0.5mm] (0,0.5) -- (0,1.0) arc (90:45:0.54cm) -- cycle;
\draw[->,line width=0.5mm] (0,0) -- (4.5,4.5);
\draw[->,line width=0.5mm] (0,2.5) -- (4.4,4.6);
\draw[->,line width=0.1mm,top color=blue,bottom color=red,middle color=green] (0,0) node[circle,fill=gray!45,draw,shading=axis,shading angle=0]{$^{4}$He} 
-- (4.5,4.5) node[circle,anchor=south west,fill=gray!45,draw,shading=axis,shading angle=110]{H};
\draw[->,line width=0.1mm] (0,2.5) node[circle,fill=gray!45,draw,ball color=red]{H$^{+}$} -- (4.4,4.6) node[]{};
\draw (2.5,2) node{$\vec{r}_{A}$} (2.5,4) node{$\vec{r}_{B}$} (2.2,2.8) node{$\vec r$} (-0.5,1.25) node{$R$} (0.4,1.3) node{$\theta$} (-0.7,0) node{$\textrm{A}$} (-0.7,2.5) node{$\textrm{B}$};
\node[label=above:--,label=right:,label=below:+,label=left:,rotate=0]{};
\end{tikzpicture}

  \caption{Definition of the Jacobi coordinates system. Red color corresponds to the positive electric charge while the blue color corresponds to the negative one. 
  Inhomogeneous colormaps inside circles illustrate the internal polarization of neutral helium and hydrogen atoms caused by the electric fields inside a triatomic configuration.}
  \label{molecule}
\end{figure}
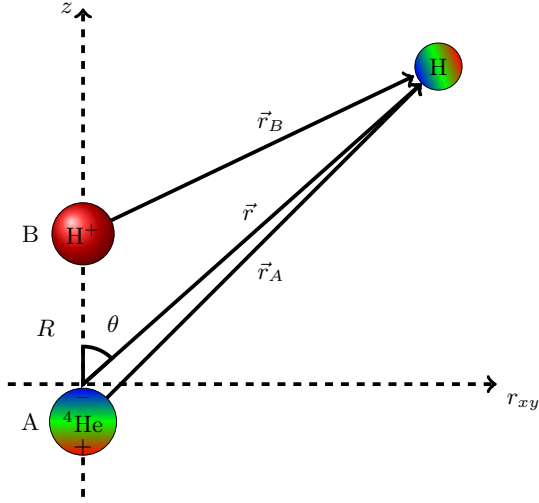

We treat the molecular ion HeH$^{+}$ as a rigid rotor and the interaction potential between the colliding partners is fully specified by two Jacobi coordinates
$V(r,\theta)$ as it is showed in Fig.~\ref{molecule}, where $r$ is the distance between the center of mass of the molecular ion and the atom H, and $\theta$ is 
the angle between the molecular symmetry axis, which coincides with the z-axis and the vector $\vec r$.  

To write down the long-range interaction potential between atomic hydrogen and helium hydride ion, we need detail information about the internal structure of HeH$^{+}$.
The quantum mechanical solutions for the wave function of the electrons at the ground and excited low-energy levels for molecular ion HeH$^{+}$
was obtained more than a half-century ago in \cite{Michels1966}. Due to his results the $^{1}\Sigma$ ground two-electron wave function of HeH$^{+}$ can be written as 
$\psi(\xi_1, \eta_1; \xi_2, \eta_2)  = \phi_A(\xi_1,\eta_1)\cdot\phi_B(\xi_2,\eta_2) + \phi_A(\xi_2,\eta_2)\cdot\phi_B(\xi_1,\eta_1)$, where 
$\phi_A(\xi, \eta) = \exp\{-\delta_A\xi-\zeta_A\eta\}$, $\phi_B(\xi, \eta) = \exp\{-\delta_B\xi-\zeta_A\eta\}$, labels 1 and 2 correspond to the first and second electrons, 
$\xi$ and $\eta$ are the prolate elliptical coordinates connected to spherical coordinates by means of the following relations (\cite{Lesiuk2014}):
$$\xi = \frac{r_A+r_B}{R},\quad \eta = \frac{r_A-r_B}{R},$$ so that $1\le\xi\le\infty$, $-1\le\eta\le 1$ and $0\le\phi\le 2\pi$, A and B denote the He and H 
nucleus in HeH$^{+}$ molecular ion, $r_A$ and $r_B$ are radial distances from these nucleus.
As it follow from \cite{Michels1966}, the minimal bound energy of HeH$^{+}$ correspond to the interatomic distance $R=0.772$~\AA, for which the approximation 
parameters are as follows $\delta_A = 1.5311143$, $\zeta_A = 1.6848429$, $\delta_B = 1.2168429$, $\zeta_B = 0.1503286$.

\begin{figure}
  \includegraphics[width=0.49\textwidth]{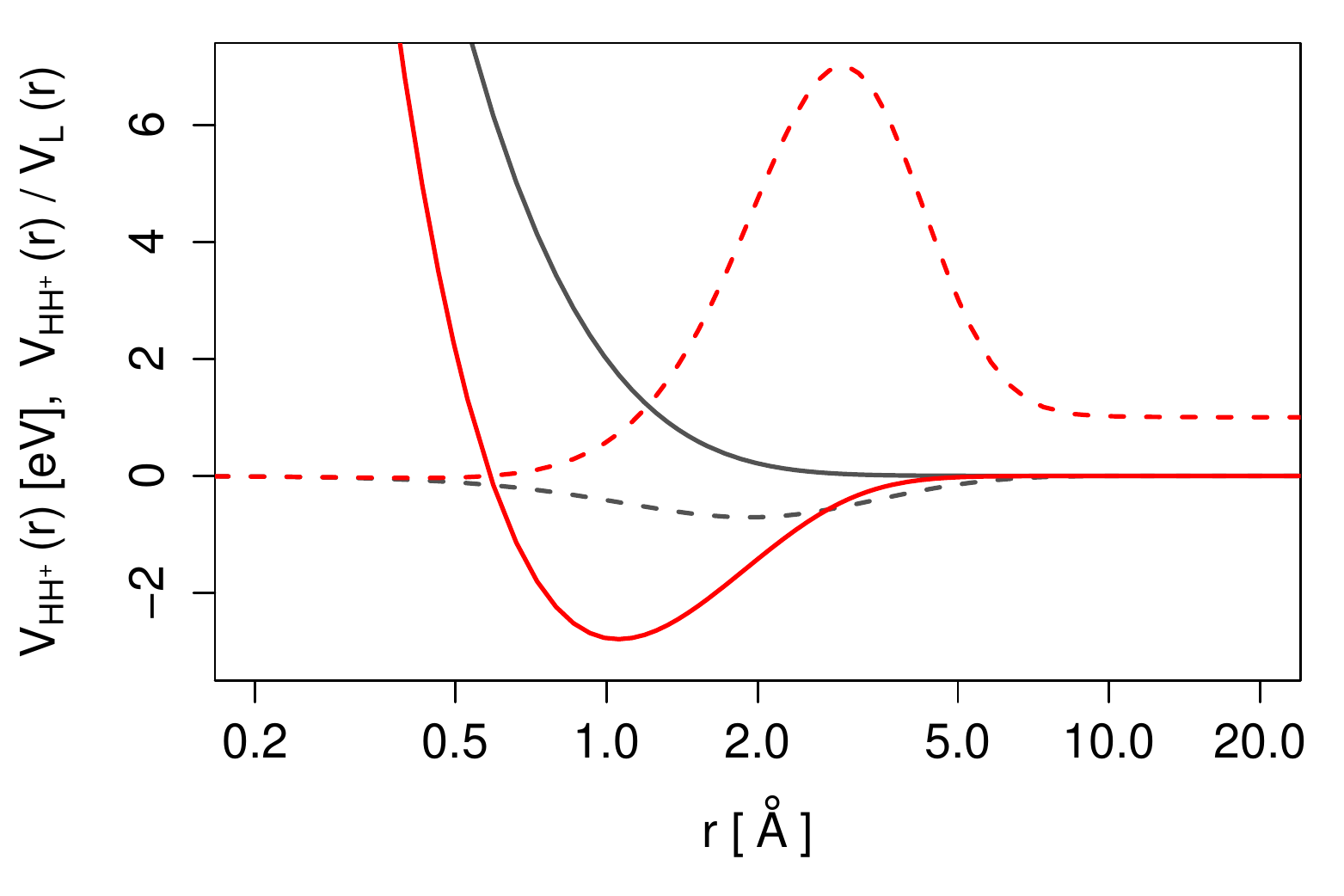}
  \caption{The two-atomic potential $V_{\rm {H-H}}^{(2)}$ from the analytical approximation proposed by \cite{Fazio2012} (solid black line) versus proton-hydrogen interaction potential $V_{\rm HH^{+}}$ \cite{Hunter1977} (solid red line). 
  Asymptotic behaviors of $V_{\rm {H-H}}$ (dashed black line) and $V_{\rm HH^{+}}^{(2)}$ (dashed red line) are shown as their relations to the long-range approximation $V_{\rm L}$.}
  \label{Pot_Hp}
\end{figure}

Since $\psi(\xi_1, \eta_1; \xi_2, \eta_2)$ is symmetric under replacement of the electrons, we can obtain the probability density for electrons spatial localization as
\begin{equation}
 \psi^2(\xi, \eta) = 2\pi\left(\frac{R}{2}\right)^3\int\limits_1^{\infty}\int\limits_{-1}^{1}|\psi(\xi, \eta; \xi', \eta')|^2 \left(\xi'^2-\eta'^2\right)d\xi' d\eta',
 \label{psi2}
\end{equation}
where we take into account that the volume element in elliptical coordinates is as follows $dV = \left({R}/{2}\right)^3\left(\xi^2-\eta^2\right)d\xi d\eta d\phi$ and
spherical coordinates are expressed through the ellipsoidal coordinates by means of the well-known expressions
\begin{eqnarray}
 && r_k = \frac{R}{2} (\xi+k\eta),\nonumber\\
 && \cos(\theta_k) = \frac{1+k\xi\eta}{\xi+k\eta},\nonumber\\
 && \sin(\theta_k) = \frac{\sqrt{(\xi^2-1)(1-\eta^2)}}{\xi+k\eta}.\nonumber
 \end{eqnarray}
where the value of $k$ is equal to +1 if the center of coordinate system is located in the center A or -1 if it is located in the center B.
It is clear that (\ref{psi2}) can be expressed as
$\psi^2(\xi, \eta) = c_1\phi_A(\xi,\eta)\cdot\phi_A(\xi,\eta) +c_2\phi_B(\xi,\eta)\cdot\phi_B(\xi,\eta) + 2c_3\phi_A(\xi,\eta)\cdot\phi_B(\xi,\eta)$,
where $c_1 = 4.995054$, $c_2 = 6.339531$ and $c_3 = 4.492389$.

In Fig.~\ref{molecule} we place the molecular ion HeH$^{+}$ along $z$ axis so that its center of mass coincide with the center of the coordinate system while 
helium and hydrogen nucleus are placed at $z_{\rm He} = - 0.2R$ and $z_{\rm H^{+}} = 0.8R$ correspondingly.
The center of the two-electron cloud is on the z-axis at
\begin{eqnarray}
 z_{2e} &=& \int\limits_{V} z\psi^2dV\\ 
 &=&  2\pi\left(\frac{R}{2}\right)^4\int\limits_1^{\infty}\int\limits_{-1}^{1}(0.6-\xi\eta)|\psi(\xi, \eta)|^2 \left(\xi^2-\eta^2\right)d\xi d\eta,\nonumber
\end{eqnarray}
where we take into account that $z=r\cos(\theta)=r_A\cos(\theta_A)+z_{\rm He}$. Integration gives $z_{2e} = -0.032R$, 
that means that the center of electrons cloud placed between helium and hydrogen nucleus and is very close to the mass center of the molecular ion.
Since the center of two-electrons cloud is only a bit shifted from 
the helium nucleus we can consider helium as a neutral atom which is polarized in the electric field of hydrogen nucleus.
The electric dipole moment of such polarized helium is equal to $\mu_{He} = 2e(z_{\rm He}-z_{2e}) \simeq - 0.259497\, e$\AA.
The total dipole moment of HeH$^{+}$ is equal to $\mu_{\rm HeH^{+}} = \mu_{\rm He} + ez_{\rm H} = 0.35810\, e$\AA, or in an atomic unit $\mu_{\rm HeH^{+}} = 0.67694\, ea_B$ 
and in a CGS unit  $\mu_{\rm HeH^{+}} = 1.72165\, \textrm{D}$.
The obtained value for dipole moment of HeH$^{+}$ coincides with the estimation obtained by \cite{Engel2005} 
(use Table.3 in that paper and make interpolation for $^4$HeH$^{+}$ at the equilibrium interatomic distance 1.455$a_B$).

So, the long-range interaction potential between atomic hydrogen and helium hydride ion can be written as follows
\begin{equation}
 V_{\rm L,H-HeH^{+}} = -\vec{\mu}_{\rm H}\vec{E}_{\rm HeH^{+}},
 \label{lrpot}
\end{equation}
where $\vec{\mu}_{\rm H} = \frac12 \alpha_{\rm H}{\vec E}_{\rm HeH^{+}}$ is the induced dipole moment of atomic hydrogen in the electric field of molecular ion HeH$^{+}$, and
$\alpha_H = (9/2)a_B^3$ is the polarizability of neutral hydrogen where $a_B$ is the Bohr radius. 
The electric field of molecular ion HeH$^{+}$ we represent as the sum of two components ${\vec E}_{\rm HeH^{+}} = {\vec E}_{\rm He} + {\vec E}_{\rm H^{+}}$ where electric fields ${\vec E}_{\rm He}$ and
${\vec E}_{\rm H^{+}}$ have next components:
\begin{eqnarray}
 {\vec E}_{He} &=& \left(\mu_{\rm He}\frac{3\cos(\theta_A)\sin(\theta_A)}{r_A^3},\mu_{\rm He}\frac{3\cos^2(\theta_A)-1}{r_A^3}\right),\\
 {\vec E}_{\rm H^{+}} &=& \left( \frac{er}{r_B^3}\sin(\theta),\frac{er}{r_B^3}(\cos(\theta)-x)\right),
\end{eqnarray}
where $r_B = r\sqrt{1+x^2-2x\cos(\theta)}$, $x = z_{\rm H}/r$, and we can use approximations $\theta_A\simeq\theta$ and $r_A\simeq r$.
 
The expression (\ref{lrpot}) is a good approximation to long-range interaction potential for $r\ge 10$~\AA. To show this let us consider the long-range 
interaction potential between proton and hydrogen $V_{\rm L,HH^{+}} = -\vec{\mu}_{\rm H}\vec{E}_{\rm H^{+}}$ well approximated at large distance $r$ by well 
known expression for charge-induced dipole interaction $V_{\rm L}(r) = -\frac12 \alpha_{\rm H} r^{-4}$.
To describe deviations of this analytical potential from the results of accurate numerical calculations for proton-hydrogen interaction performed by \cite{Hunter1977} 
(see Table.~1 of that paper) we introduce the screening function
 \begin{equation}
  F(r) = (C_{00}(r)-C_{00}(\infty))/V_{\rm L}(r),
  \label{Fr}
 \end{equation}
that, as it is shown by the red dashed line in Fig.~\ref{Pot_Hp}, at distances $r < 10$ \AA { } deviates from unity. 
Therefore we have to modify approximation for potential interaction between 
proton and hydrogen in (\ref{lrpot}) as follows $V_{\rm HH^{+}}(r,\theta) = F(r_B(r,\theta))\cdot V_{\rm L,HH^{+}}(r,\theta)$. 
\begin{figure}
  \includegraphics[width=0.49\textwidth]{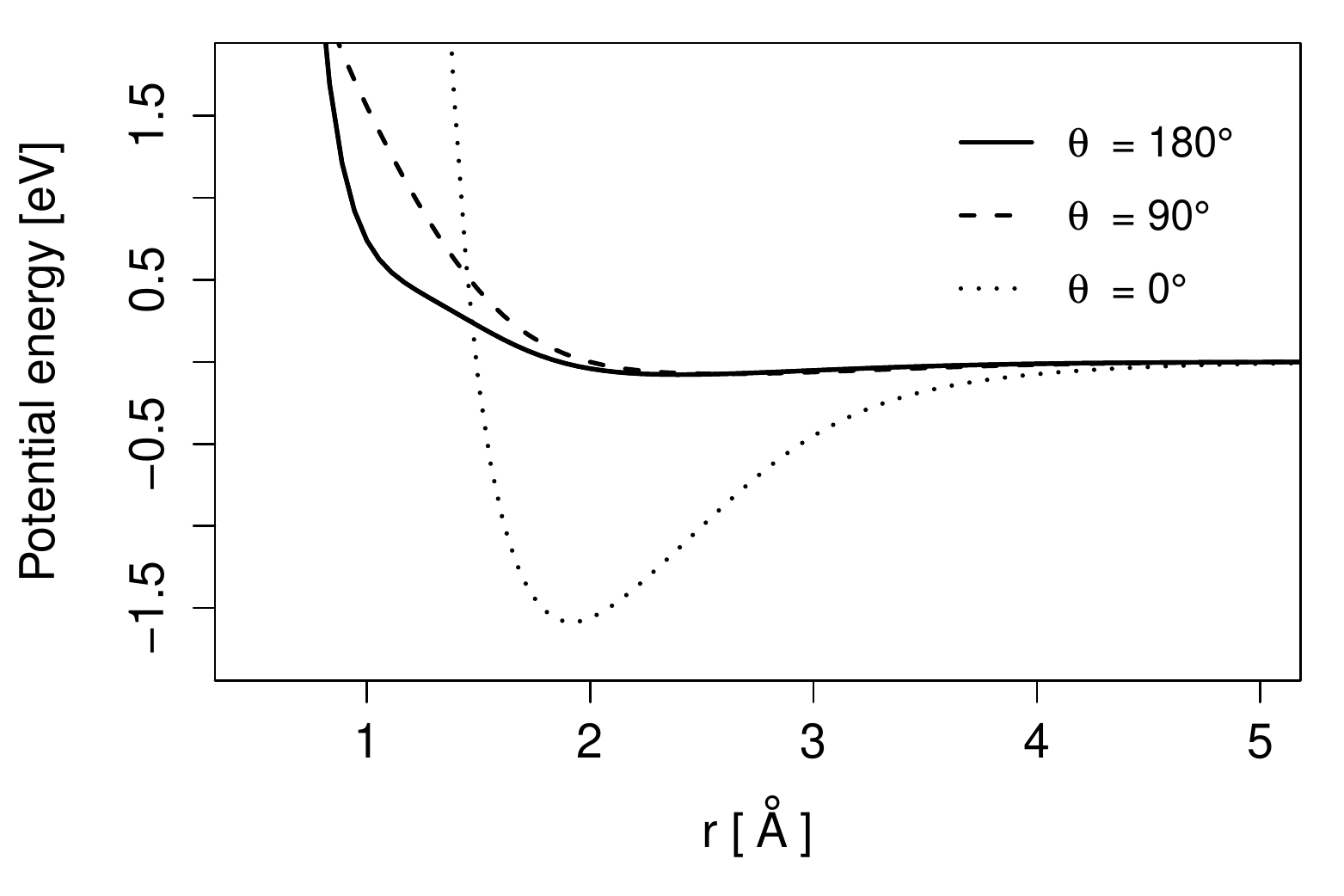}
%  \plotone{Potential_prof.pdf}
  \caption{Cuts through PESs at different angles: $\theta = 0^{o}$ -- dotted line, $\theta = 90^{o}$ -- dashed line, and $\theta = 180^{o}$ -- solid line with HeH$^{+}$(R) + H as the zero of energy.}
  \label{Potential_prof}
\end{figure}

\begin{figure}
%  \begin{center}
\includegraphics[width=0.49\textwidth]{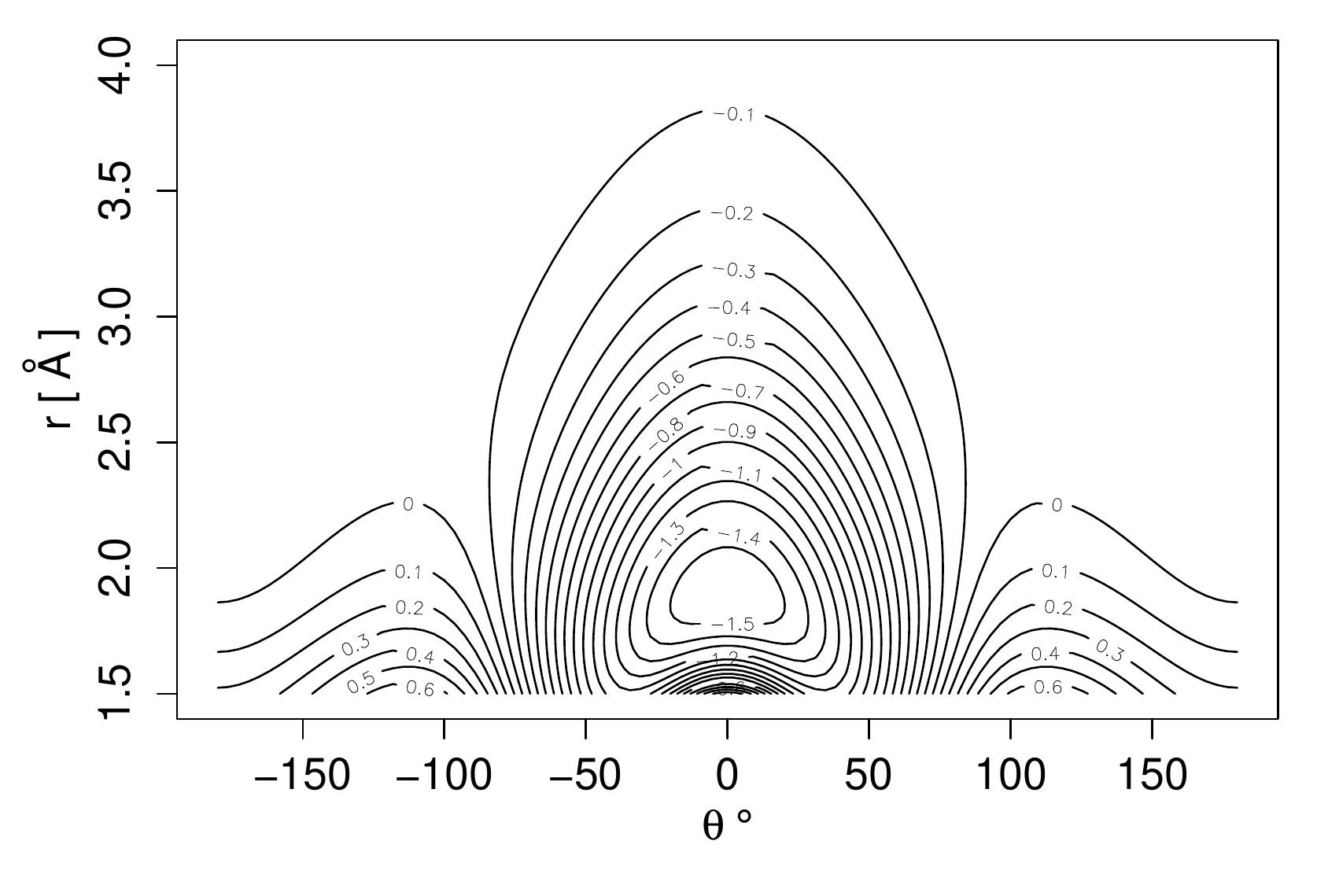}
 %\end{center}
  \caption{
   PESs $V(r,\theta)$ (in eV) considered for H--HeH$^{+}$, where $r$ (in \AA) represents the distance between the 
  atom  and  the  center of mass  of  the  molecular  ion,  whereas $\theta$ (in degrees) is the relative angle between $r$ and the 
  orientation of the molecular ion HeH$^{+}$.}
  \label{molprobv}
\end{figure}

Since the deviation of the screening function (\ref{Fr}) from unity is a result of the internal re-ordering of charges within atomic hydrogen by the predominantly inhomogeneous electric field of proton, we need to take this into account also for the dipole interaction of hydrogen and helium, i.e. $V_{\rm HHe}(r,\theta) = F(r_B(r,\theta))V_{\rm L,HHe}(r,\theta)$, 
where $V_{\rm L,HHe} = -\vec{\mu}_{\rm H}\vec{E}_{\rm He}$ is the long-range interaction between polarized hydrogen and helium.
Thus, the modified long-range interaction between atomic hydrogen and the helium hydride ion take a form
$V_{\rm H-HeH^{+}}(r,\theta)\simeq F(r_B(r,\theta))V_{\rm L,H-HeH^{+}}(r,\theta)$, where $V_{\rm L,H-HeH^{+}}(r,\theta) = V_{\rm L,H-H^{+}}(r,\theta)+V_{\rm L,H-He}(r,\theta)$.

The one of the latest version of the short-range analytical approximation, PM10 \cite{Fazio2012}, for the three atomic PESs considered for HeH$_2^{+}$  
\begin{eqnarray}
 V(R,r_A,r_B) &=& V^{(2)}_{{\rm H-He}}(R) + V^{(2)}_{{\rm H-He}}(r_A)\nonumber\\ 
 &+& V^{(2)}_{{\rm H-H}}(r_B)+ V^{(3)}_{{\rm H-He-H}}(R,r_A,r_B),
 \label{threatomicpot}
\end{eqnarray}
designed to describe He + H$_2^{+}$~$\to$ HeH$^{+}$ + H reaction and considered more accurate than previous ones (\cite{Joseph1987,Aguado1992,Aquilanti2000,Palmieri2000,Xu2008,Ramachandran2009}) is used to describe the interaction potential between atomic hydrogen and helium hydride ion at interatomic distances not exceeding $\sim 5$~\AA{}. 
The threshold function $\Pi(r)=\left(1+\tanh\left[2\cdot(r-{4.5\,\textrm{\AA}})\right]\right)/2$ is used to switch between the short-range two-atomic potentials $V^{(2)}_{{\rm H-H}}$ and $V^{(2)}_{{\rm H-He}}$, that are exponentially suppressed at $r_B,\, r_A \sim 5$~\AA{}, and the long-range two-atomic potentials $V_{\rm L,H-H^{+}}(r,\theta)$ and $V_{\rm L,H-He}(r,\theta)$ in the following way: $V^{(2)}_{\rm H-He}(r_A(r,\theta))\times \left[1-\Pi (r_A(r,\theta))\right] + V_{\rm L,H-He}(r,\theta)\times \Pi (r_A(r,\theta))$ and $V^{(2)}_{\rm H-H}(r_B(r,\theta))\times \left[1-\Pi (r_B(r,\theta))\right] + V_{\rm L,H-H^{+}}(r,\theta)\times \Pi (r_B(r,\theta))$ correspondingly.

In Fig.~\ref{molprobv} the resulting PES considered for H--HeH$^{+}$ interaction is shown as a contour map in the plane of two Jacobi coordinates, $r$, and $\theta$. A 1.59 eV deep van der Waals well is placed at approximately $r\simeq 1.93$~\AA\, and $\theta = 0^{o}$ correspond to the He-H-H configuration of HeH$_2^{+}$ system with HeH$^{+}$(R) + H as the zero of energy (see Fig.~\ref{Potential_prof}). Potential has a repulsive barriers at $r\simeq 1$~\AA\ for $\theta = 90^{o}$ and  $\theta = 180^{o}$, and at $r\simeq 1.5$~\AA\ for $\theta = 0^{o}$.

\section{State-to-state integral cross sections and rate coefficients} \label{sec:floats}

\begin{figure}
  \includegraphics[width=0.49\textwidth]{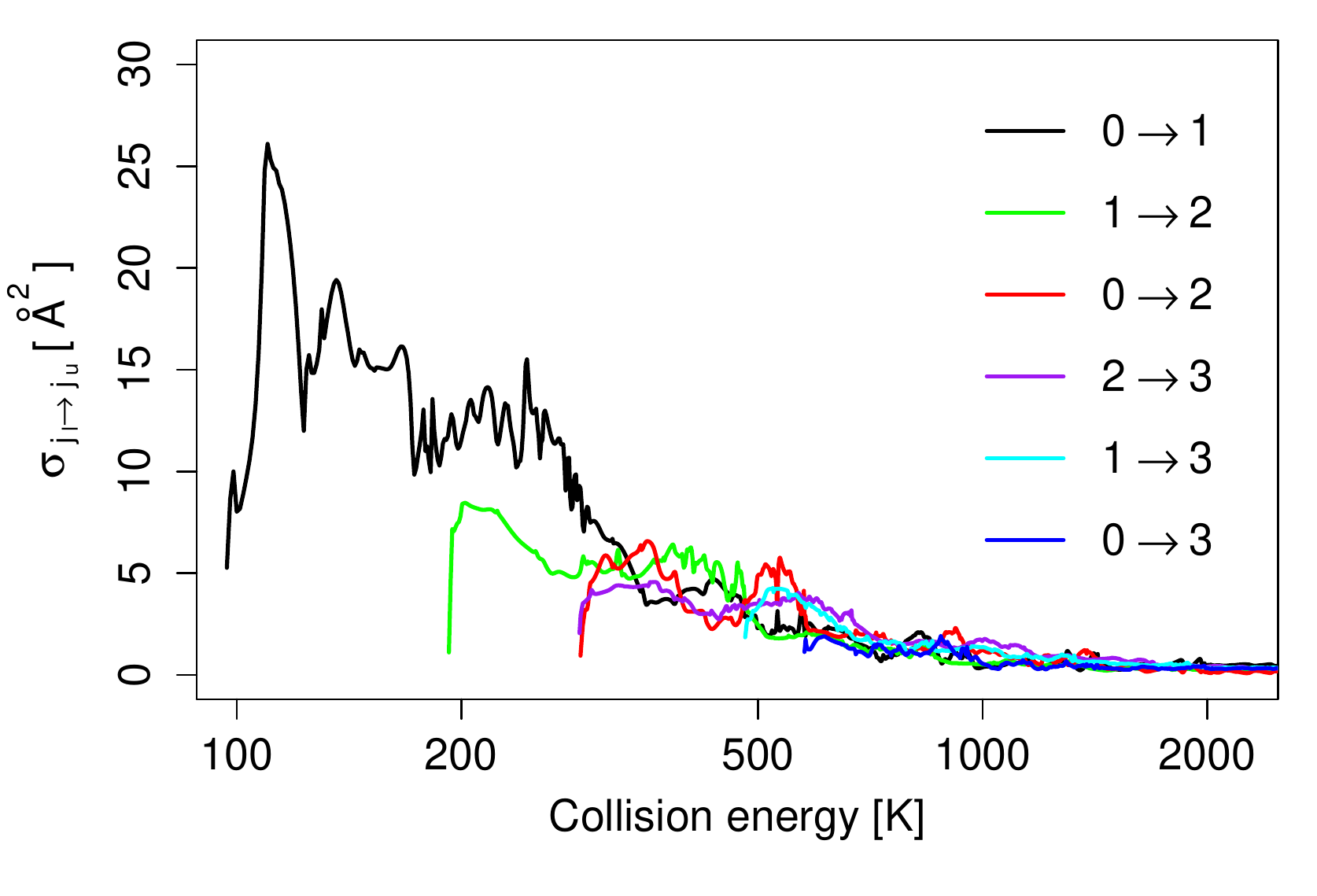}
  \caption{Cross-section of inelastic H--HeH$^{+}$ scattering for rotational transitions in HeH$^{+}$ molecular ion as a function of collision energy.}
  \label{Sigma}
\end{figure}

The rovibrational energy structure of $^4$HeH$^{+}$ (i.e. HeH$^{+}$) and its isotopologues $^3$HeH$^{+}$, $^3$HeD$^{+}$, and $^4$HeD$^{+}$, were calculated previously in \cite{Engel2005}  (see also \cite{Amaral2019}) and results
are available with corresponding Einstein coefficients at \href{http://www.exomol.com/data/molecules/HeH_p/}{www.exomol.com}. In Table~\ref{Levels} we listed this data for six lowest rotational levels within the vibrational ground state of HeH$^{+}$.
The MOLSCAT code (ver.14, \cite{Molscat}) was used to calculate 
state-to-state cross sections for inelastic scattering of hydrogen atoms on HeH$^+$ molecular ions.
The calculated inelastic scattering cross-section for low-levels rotational transitions in HeH$^{+}$ ion as a function of collision energy is shown in Fig.~\ref{Sigma}.

\begin{table}[h!]
\caption{The lovest rotational energy levels data for HeH$^{+}$. }
\label{Levels}
\centering
%\begin{ruledtabular}
\begin{tabular}{cccc}
\hline\hline
Frequency $\nu_{ul}$ & Transitions & $E_u$ & $A_{ul}$  \\
{} [GHz]  & $j_u$ - $j_{l}$ & [K] & [$s^{-1}$]  \\
\hline
 2010 & 1 -- 0 & 96 & 0.109  \\
 4009 & 2 -- 1 & 289 & 1.04  \\
 5984 & 3 -- 2 & 576 & 3.75  \\
 7925 & 4 -- 3 & 956 & 9.14  \\
 9821 & 5 -- 4 & 1428 & 18.1 \\
\hline
\end{tabular}
%\end{ruledtabular}
\label{Tab1}
\end{table}

\begin{figure}
  \includegraphics[width=0.49\textwidth]{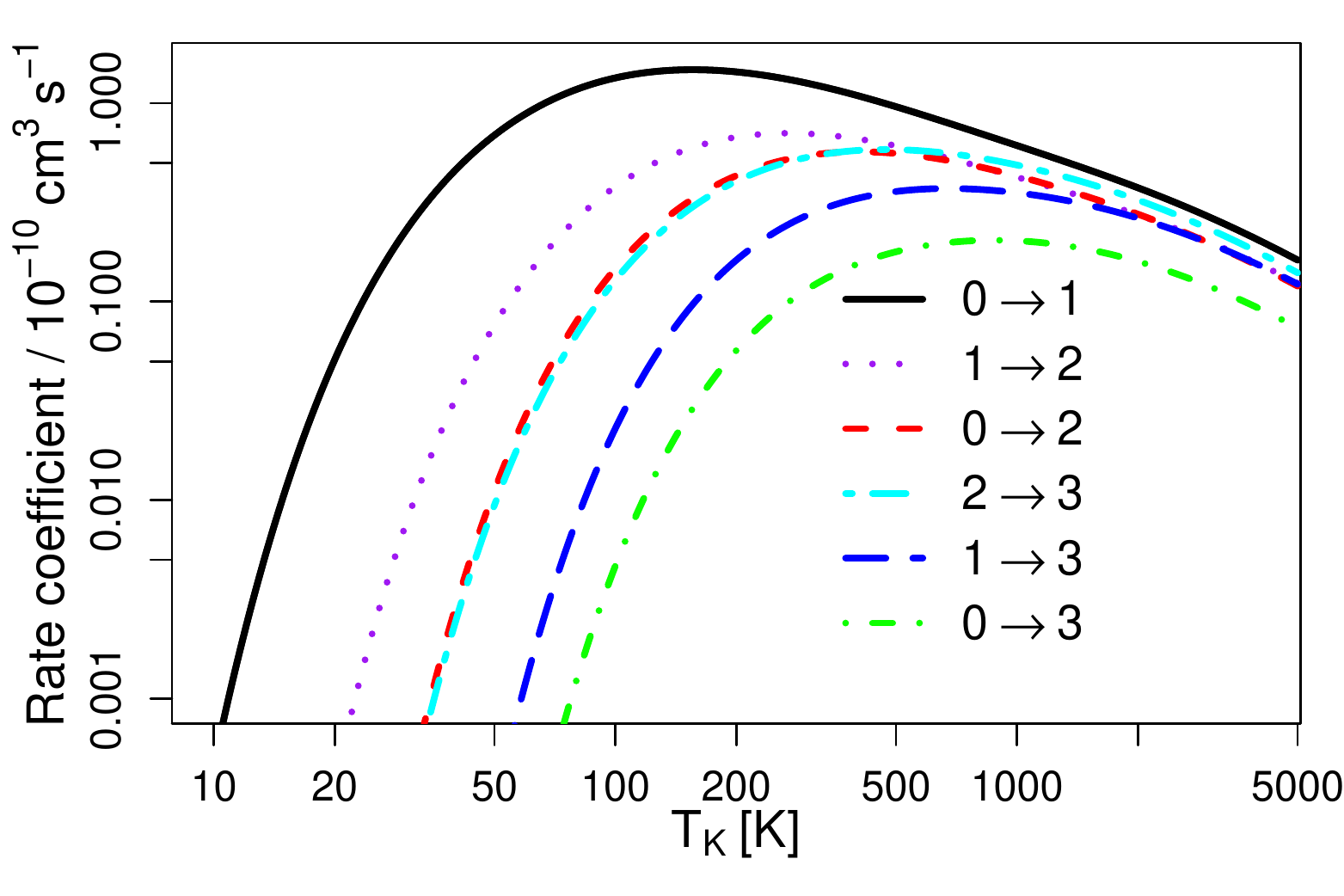}
  \includegraphics[width=0.49\textwidth]{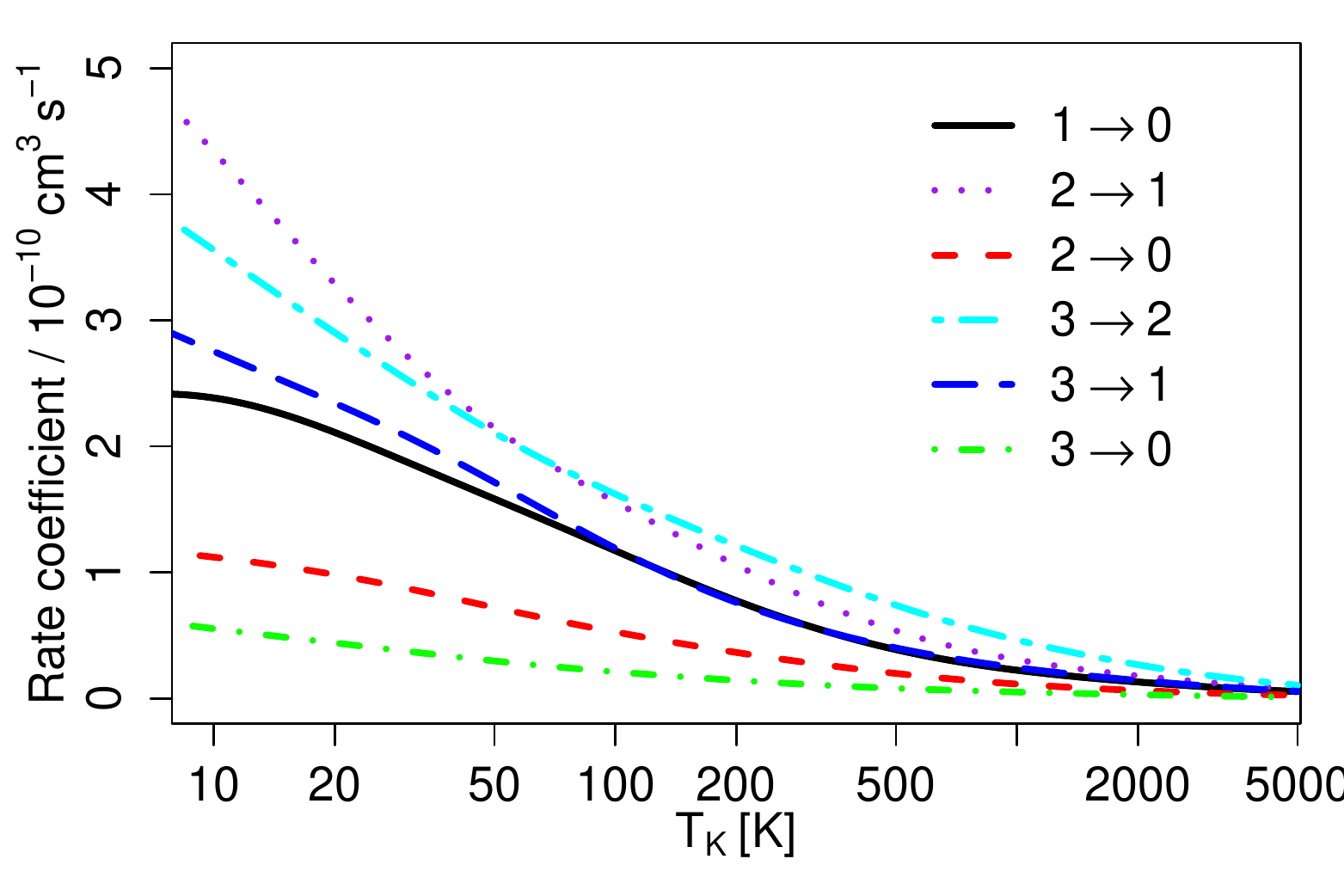}
  \caption{The rate coefficients of collisional excitation (top panel) and de-excitation (bottom panel) of HeH$^+$ rotational levels by H at different gas temperatures.}
  \label{Rate_exitation}
\end{figure}

By assuming a Maxwell--Boltzmann distribution over the collision energy, we calculated the state selected de-excitation rate coefficients as follows
\begin{eqnarray}
 k_{u\to l}(T) &=& \left[\frac{8}{\pi\mu(k_BT)^3}\right]^{1/2}\exp\left(\epsilon_u/k_BT\right)\times \nonumber\\
 && \int\limits^{\infty}_{0}\sigma_{u\to l}(E)(E-\epsilon_u)\exp\left(-E/k_BT\right)dE,
\end{eqnarray}
where $\epsilon_u$ is the energy of the $u$-th rotational level, $\mu$ is the reduced mass and $E$ is the total energy equal to the sum of collision energy and energy of the initial state of the molecular ion, $\epsilon_u$.  The results of the calculations are shown on the upper panel in Fig.~\ref{Rate_exitation}. The reverse transition rate (excitation) coefficients can be obtained through the expression
\begin{equation}
 k_{l\to u}(T) =  \frac{2j_u+1}{2j_l+1}\exp\left\{(\epsilon_u-\epsilon_l)/k_BT\right\}k_{u\to l}(T).
 \label{kul_to_klu}
\end{equation}
The dependences of excitation rate coefficients on kinetic temperature are shown in Fig.~\ref{Rate_exitation} (lower panel).
We fitted the calculated de-excitation rate coefficients by the simple formula 
\begin{equation}
k_{u \to l}^{fit}(T)=10^{-10} \sum\limits_{i=0}^{5} a_i x^i \quad [\textrm{cm}^3/\textrm{s}], 
 \label{k_ul_approximation}
\end{equation}
where $x = \log_{10}(T)$ and dimensionless approximation coefficients $a_0$, $a_1$, $a_2$, $a_3$, $a_4$ and $a_5$ for a few lowest rotational transitions are listed in Table.~\ref{Approx_coef} in Appendix.
The excitation rate coefficients can be easy evaluated from the approximations for de-excitation rate coefficients (\ref{k_ul_approximation}) through formula (\ref{kul_to_klu}), 
where values $(\epsilon_u - \epsilon_l)/k_B$ for corresponding $j_u$ and $j_l$  are listed in Table.~\ref{Approx_coef}.


\section{$\textrm{HeH}^{+}$ collisional (de)excitation in Dark Ages} \label{sec:floats}

\begin{figure}
  \includegraphics[width=0.49\textwidth]{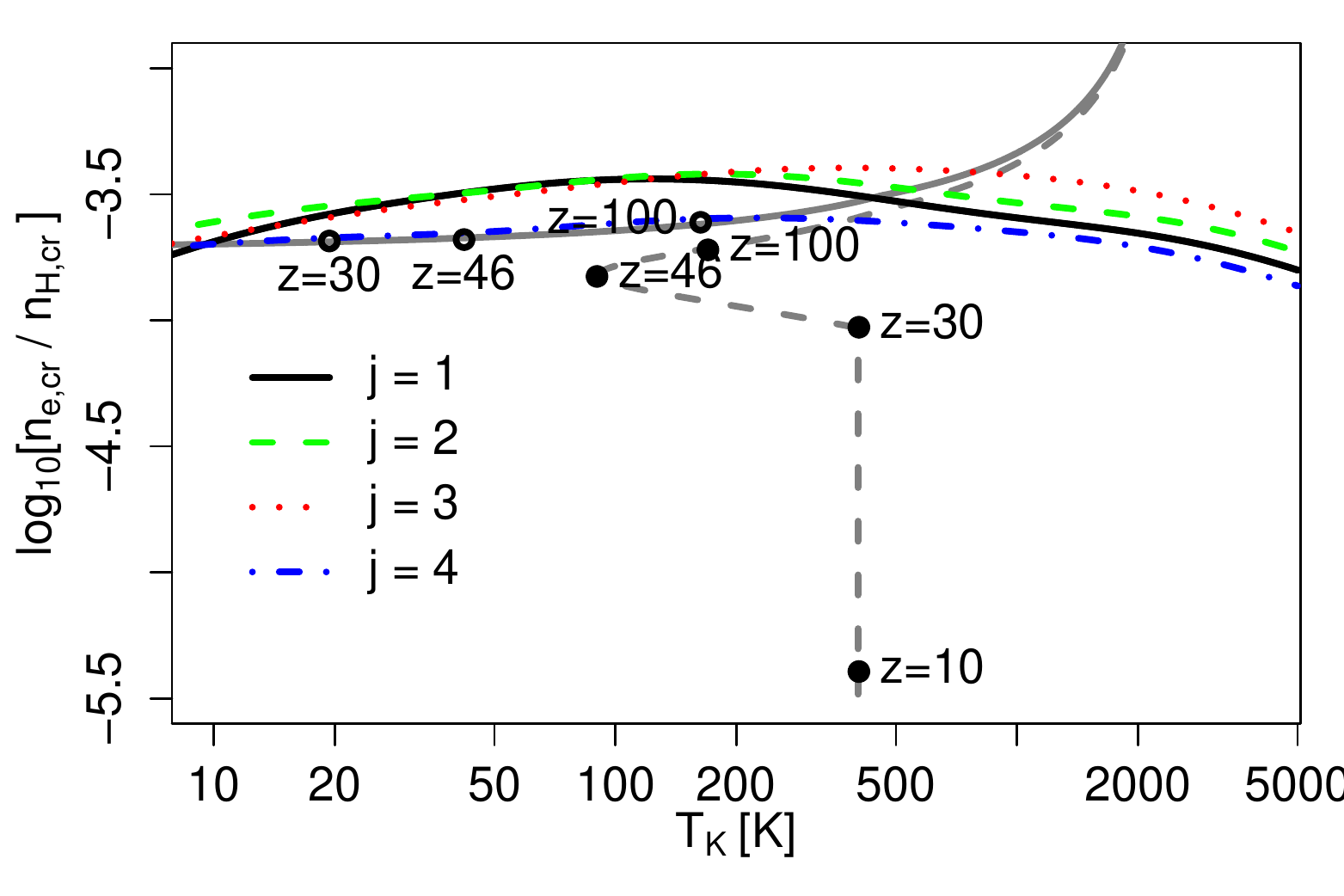}
  \caption{The ratios of critical densities $n_{\rm e,cr}/n_{\rm HI,cr}$ for rotational levels with $j=1-4$ and ratio $n_{\rm e}/n_{\rm HI}$ for different gas temperatures in Dark Ages. The solid gray line depicts the ratio $n_{\rm e}/n_{\rm HI}$ on the cosmological background, the dashed gray line inside the halo which is virialized at $z\simeq 30$.}
  \label{cr_ionization}
\end{figure}

To understand the role of collisional de-excitations of HeH$^{+}$ by atomic hydrogen we need to compare them with collisional 
de-excitations this molecular ion by free electrons as well as with spontaneous emission. For this in Table.~\ref{critical_densities} in Appendix we present the estimation of the critical densities in cm$^{-3}$ for collisional e$^{-}$--HeH$^{+}$ and H--HeH$^{+}$ de-excitations which are defined for a rotational level $j$ as follows
\begin{eqnarray}
  n_{\rm e,cr}(j) &=& \sum\limits_{j'<j} \frac{A(j\to j')}{k_{\rm e}(j\to j')},\\
  n_{\rm HI,cr}(j) &=& \sum\limits_{j'<j} \frac{A(j\to j')}{k_{\rm H}(j\to j')},
\end{eqnarray}
where we can assume that the Einstein $A$-coefficients are dominated for $\Delta j = 1$ transitions, the rate coefficients for rotational transitions induced by collisions with free electrons, for $k_{\rm e}(j\to j')$ we used the rates given by \cite{Khamesian2018}. Since, the rates of collisional de-excitations by electrons and atomic hydrogen are proportional to their number densities,
\begin{equation}
C^{\rm e}_{ij}\equiv n_{\rm e} k_{\rm{e},ij}, \quad C^{\rm H}_{ij}\equiv n_{\rm HI} k_{{\rm H},ij},\nonumber 
\end{equation}
one can compare their values by ratio
\begin{equation}
\frac{C^{\rm e}_{ij}}{C^{\rm H}_{ij}}=\frac{n_{\rm HI,cr}(j)}{n_{\rm e,cr}(j)}\frac{n_{\rm e}}{n_{\rm HI}}.\nonumber
\end{equation}
The last term in the right hand side of expression in the conditions of dark ages halos is close to the value of degree of hydrogen ionization or electron-proton fraction $x_{\rm p}\approx x_{\rm e}\equiv n_{\rm e}/(n_{\rm HI}+n_{\rm HII})$. In Figure~\ref{cr_ionization} we present the ratios of critical densities $n_{\rm e,cr}/n_{\rm HI,cr}$ for rotational levels with $j=1-4$ and ratio $n_{\rm e}/n_{\rm HI}$ for different gas temperatures in the Dark Ages and in the halo which is virialized at $z=30$. One can see that ratio $n_{\rm e}/n_{\rm HI}$ in Dark Ages at $z<100$ is a bit lower than ratios $n_{\rm e,cr}/n_{\rm HI,cr}$ for levels with $j=1,2,3$, and close to this ratio for level with $j=4$. It means that collisional excitation/de-excitation of the lowest rotational levels of molecular ion HeH$^+$ by electrons and atomic hydrogen are comparable and both must be taken into account.

The fraction of electrons in the halo decreases over time due to their recombination since the process of electron recombination is stronger in denser regions and less for areas with lower gas concentration \cite{Novosyadlyj2006}. So, de-excitations of the HeH$^{+}$ rotational levels through collisions with atomic hydrogen dominate over the de-excitations through collisions with electrons in more dense regions such as halos. As an example, the filled circles in Fig.~\ref{cr_ionization} show the evolution of ratio $n_{\rm e}/n_{\rm HI}$ in the overdensity region  
that reaches of virialization at $z=30$. One can see, that number density of electrons decreases after virialization despite the constant gas density (see for details Fig.~4 in \cite{Novosyadlyj2018}). As a result, at $z=10$  de-excitations of the HeH$^{+}$ rotational levels through collisions with atomic hydrogen start to dominate over the de-excitations in the electron-HeH$^{+}$ inelastic collisions by two orders of magnitude.

\section{Emission of dark ages halos in the rotational lines of $\textrm{HeH}^{+}$} \label{sec:cite}

\begin{table*}[!ht]
\begin{center}
\caption{Physical values and chemical composition of halos with mass $M_h=5.3\cdot10^9$ M$_{\odot}$ and different amplitude of initial curvature perturbation  $C_k$ (seed of halo the wave number of corresponding linear perturbation is $k=5.0$ Mpc$^{-1}$), which are virialized at $z_v$\footnote{$\rho_{m}$ is the matter density virialized halo, $T_K$ is kinetic temperature of baryonic gas, $n_{\rm HI}$ is the number density of neutral hydrogen atoms, $n_{\rm p},\,n_{\rm e}$ are the number densities of protons and electrons at $z=z_v/z=10$, $n_{\rm HeH^{+}}$ is the number densities of molecular ion HeH$^{+}$, $r_h$ is the radius of halo in comoving coordinates, $\theta_h$ is the angular radius of geometrically limited halo.}.}
\begin{tabular} {cccccccccccc}
\hline
\hline
   \noalign{\smallskip}
$C_k$&$z_v$&$\rho_{m}$&$T_K$&$n_{\rm HI}$&$n_{\rm p}\approx n_{\rm e}$&$n_{\rm HeH^{+}}$&$r_h$&$\theta_h$\\
 \noalign{\smallskip} 
[$10^{-4}$]& &[$10^{-24}$g/cm$^3$]&[K]&[cm$^{-3}$]&[10$^{-6}$cm$^{-3}$]&[10$^{-15}$cm$^{-3}$]&[kpc]&[arcsec]\\
 \noalign{\smallskip} 
\hline
   \noalign{\smallskip} 
3.0 &30.41&15.20&402.1&1.14&106.2/3.8&3.65&1.78&1.03\\ 
2.5 &25.15&8.79&298.9&0.66& 66.1/4.0&2.63&2.14&1.05\\ 
2.0 &19.90&4.49&206.3&0.34& 36.7/4.0&1.76&2.68&1.09\\ 
1.5 &14.65&1.89&124.3&0.14& 17.0/4.4&1.05&3.60&1.15\\  
1.0 & 9.41&0.56& 59.8&0.04&  5.6/5.0&0.50&5.38&1.26\\    
  \hline
\end{tabular}
\label{Tab4a}
\end{center}
\end{table*}

Dark ages hide the mystery of the formation of the first stars in the Universe, also known as the first-generation stars or Population III stars. 
The formation of stars is known occurs during the gravitational compression of protostar gas clouds, which can cool by converting the kinetic energy of atoms and molecules into electromagnetic radiation in the processes of their inelastic scattering.
Gas cooling is inefficient when the energy of collisions between atoms and molecules becomes less than their lowest excitation threshold. The subsequent collapse of the protostar cloud would continue if its mass, thanks cooling, will be larger than the Jeans mass, which depends on the gas temperature as $\sim T^{3/2}$.
Thus, the lowest energy of excitation of gas ingredients is connected with the lower mass bound of the forming stars.
The lowest excitation energies correspond to lowest rotational and vibrational transitions of the molecules as well as the excitation of the external electrons in the heavy multi-electron atoms. However, gas in the Dark Ages does not contain heavy elements, since they were synthesized somewhat later in the nuclei of the first stars or during their supernova bursts.
Therefore, only small-mass atomic molecules such as H$_2$, HD, HeH$^{+}$, LiH, etc. were involved in the processes of gas cooling in the Dark Ages.
Despite the small abundance of these molecules in the early Universe \cite{Bovino2011,Novosyadlyj2017,Novosyadlyj2018}, they played a crucial role in the formation of the first stars. Therefore, observation of these molecules in the Dark Ages is one of the most important tasks of modern cosmology. Below, we estimate the differential brightness temperatures caused by thermal collisions and resonant scattering of CMB quanta in the rotational lines of HeH$^{+}$ ions.

\subsection{Halo formation and chemistry in the Dark Ages}

The set of the physical conditions and chemistry of the halos in the Dark Ages we obtain by modeling the evolution of individual spherical perturbations in the four-component Universe (cold dark matter, baryon matter, dark energy, and thermal relict radiation) by integrating the appropriate system of equations described in \cite{Novosyadlyj2016,Novosyadlyj2018}. 
The complete set of the cosmological parameters used in the paper is as follows: the Hubble constant $H_0=70$ km/s/Mpc, the mean density of baryonic matter in the units of critical one $\Omega_b=0.05$, the mean density of dark matter $\Omega_m=0.25$, the mean density of dark energy $\Omega_{de}=0.7$, its equation of state parameter $w_{de}=-0.9$, the effective sound speed $c_s=1$ (in units of speed of light). 
All physical values and chemical composition of a halo with mass $M_h=5.3\cdot10^9$ M$_\odot$, which are necessary for computation of the excitations and brightness temperatures in the molecular rotational lines, are presented in Tab.~\ref{Tab4a}. The data for halos of other masses, $6.6\cdot10^8$, $8.3\cdot10^7$, $1.0\cdot10^7$ and $1.3\cdot10^6$M$_\odot$, are presented in Tab.~\ref{Tab4b} in Appendix.

For simplicity, we assume that halos are homogeneous top-hat spheres forming from the primordial cosmological density perturbations due to their gravitational instability at the moment of reaching the virial density $\rho_m^{vir}(z_v)=\Delta_v\bar{\rho}_m(z_v)$, where $z_v$ is the redshift of halo virialization and we assume $\Delta_v=178$. Since this moment, halos have treated as the static gravity-bound systems with constant densities, $\rho_m^{vir}(z\le z_v)=\rho_m^{vir}(z_v)$, where chemical kinetic continues.
The mass of each halo $M_h$ in the solar mass, its radius in comoving coordinates $r_h$ [kpc] and the wave number $k$ [Mpc$^{-1}$] of initial perturbation from which halo is formed, are connected by relations \cite{Novosyadlyj2019}
\begin{equation}
\frac{M_h}{M_{\odot}}=1159\Delta_v(1+z_v)^3\Omega_mh^2r_h^3=4.5\cdot10^{12}\Omega_mh^2k^{-3},\nonumber
\end{equation}
where  $h\equiv H_0/100\rm{km/s/Mpc}$.

Since processes of virialization can take some time after virial density reached, we assume the adiabatic temperature of the gas inside newly formed haloes. 
In this case, according to \cite{Novosyadlyj2018}, the number density of HeH$^{+}$ inside virialized halos decreases with time due to a higher destruction rate of these molecular ions in a more dense medium. For halos with higher final temperatures of baryonic matter, the virial ones, the number density of HeH$^{+}$ ions vanishes via dominating of destruction collisions \cite{Novosyadlyj2018}.
Let us note that regardless of the assumption about the final baryonic gas temperature, there would be a peak-like time luminosity of halos in lines of molecular ion HeH$^{+}$ with the maximum at the moment of virialization when the number density of HeH$^{+}$ reached his maximum.

\subsection{Population of $\textrm{HeH}^{+}$ excited states}

The kinetic equations for populations of rotational levels of HeH$^{+}$ ions
\begin{equation}
 \frac{dn_i}{dt} = \sum\limits_{j\ne i}n_jR_{ji} - n_i\sum\limits_{j\ne i}R_{ij},
 \label{kinetic_eqs}
\end{equation}
where indexes $i$ and $j$ mark rotational levels, $n_i$ -- are number densities of molecules with state $i$,  and $R_{ij}$ -- are rate coefficients for up/down-levels transitions.

\begin{figure*}
\includegraphics[width=\textwidth]{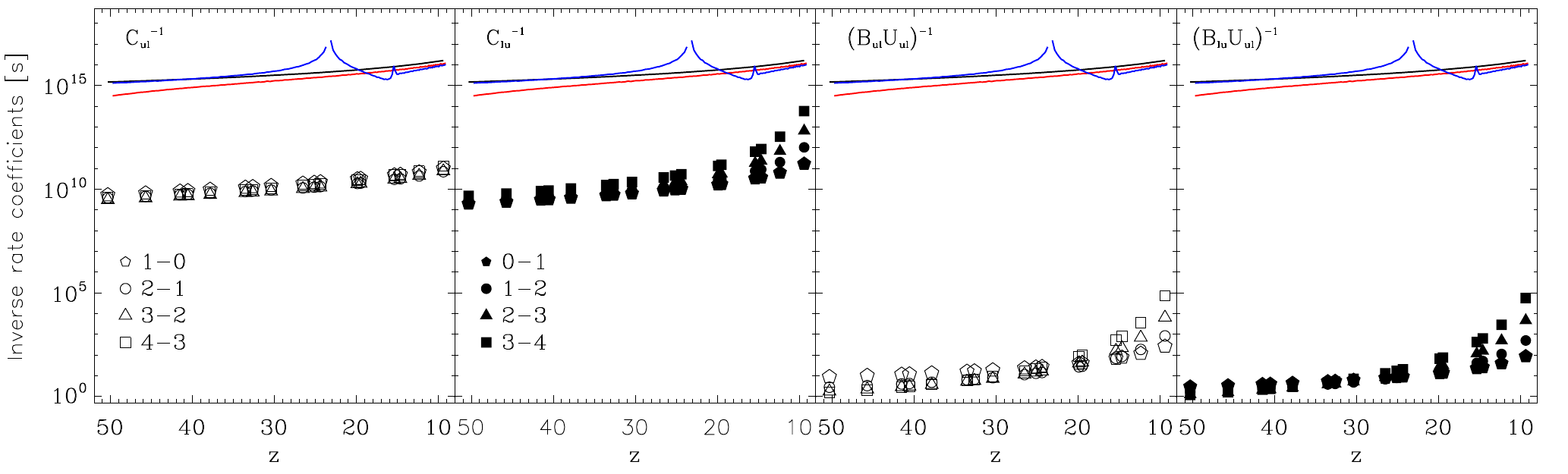}
\caption{The inverse rate coefficients $C_{ul}^{-1}$, $C_{lu}^{-1}$, $(B_{ul}U_{\nu_{ul}})^{-1}$ and $(B_{lu}U_{\nu_{ul}})^{-1}$ for molecular ion HeH$^+$ (bottom panel) in halos virialized at different $z$. 
The black solid line shows the age of the Universe corresponding to $z$, 
the red solid line shows the character time of number density change of molecular ion HeH$^{+}$, i.e. $(d\ln(n_{HeH^{+}})/dt)^{-1}$,
in halo virialized at $z\approx50$ and blue line at $z\approx15$.}
\label{inv_rates}
\end{figure*}

The number densities of e$^{-}$ and H species in halos formed in Dark Ages (see Tab.~\ref{Tab4a}  and Tab.~\ref{Tab4a}) are much lower than their critical values given in Table.~\ref{critical_densities}.
By that means that the frequencies of e$^{-}$--HeH$^{+}$ and H--HeH$^{+}$ scattering excitation/de-excitations, $C^e_{ij}$ and $C^H_{ij}$, are much lower than the 
frequencies of CMB relic photon-HeH$^{+}$ excitations, $B_{lu}U_{\nu_{ul}}$, and spontaneous emission, $A_{ul}$. 
Besides, as can be seen from Fig.~\ref{inv_rates}, all these values ​​are essentially larger than the characteristic times of changes in the rate of expansion of the Universe and 
concentration of molecular ions HeH$^{+}$ within considered range of redshifts. Consequently, the values in the left-hand side of (\ref{kinetic_eqs}) can be ignored in comparison with values 
on the right side of these equations. So, to obtain the number densities of molecules in the ground and excited rotational states we can use the following system of the linear equations:
\begin{equation}
 n_i\sum\limits_{j\ne i}R_{ij} = \sum\limits_{j\ne i}n_jR_{ji},\quad
 \sum\limits_{j}n_j=n_{HeH^{+}},
 \label{system_ni}
\end{equation}
where $R_{ij} = A_{ij} + B_{ij}U_{\nu_{ij}} + C_{ij}$, $A_{ij}$ are not equal to zero only for transitions between adjacent levels from top to botto and $C_{ij} = k^H_{ij}n_H+k^e_{ij}n_e$.
The occupations of the rotational levels can be expressed in the terms of excitation temperatures as follows
$$T_{ex,ul}=\frac{h\nu_{ul}}{k_B}\left[\ln{\frac{g_un_l}{g_ln_u}}\right]^{-1},$$
where $u=l+1$. 

\subsection{The brightness temperatures of dark ages halos}

The optical depth of spherical homogeneous top-hat halo for rotational $\nu_{ul}$ line is as follows \cite{Maoli1996,Novosyadlyj2019}
\begin{equation}
\tau_{ul}=1.55\cdot10^{50}n_l\frac{g_u}{g_l}\frac{A_{ul}}{\nu_{ul}^3}\sqrt{\frac{m_A}{T}}\left[1-\exp{\left(-\frac{h\nu_{ul}}{k_B T_{ex}}\right)}\right]r_h,
\end{equation}
where $m_A$ is the mass of HeH$^{+}$ molecular ion in unified atomic mass units, and $r_h$ is the radius of the halo with mass $M_h$ which can be estimated as follows
$$r_h = 7\cdot \left(\frac{M_h}{10^7\, M_{\odot}}\right)^{1/3}\frac{1}{1+z_v}\, \textrm{kpc},$$
where $z_v$ is the moment of halo virialization.
Results of computations $\tau$ for halos with mass $M_h=5.3\cdot10^9$ M$_\odot$ are presented in Tab.~\ref{Tab5}, for other masses in Tab.~\ref{Tab8} in Appendix.

\begin{table}[h!]
\begin{center}
\caption{The opacity of dark ages halos with mass $M=5.3\cdot 10^{9}$~$M_{\odot}$ in rotational lines of molecular ion HeH$^{+}$ formed at different redshifts $z=10-30$. Powers of 10 are given in parentheses.}
\begin{tabular} {cccccc}
\hline
\hline
   \noalign{\smallskip}
$z$& & &$\tau$&  \\
 \noalign{\smallskip} 
&0-1&1-2&2-3&3-4\\
 \noalign{\smallskip} 
\hline
   \noalign{\smallskip} 
30.41&0.949(-06) & 0.815(-06) & 0.141(-06) & 0.681(-08) \\ 
25.15&0.121(-05) & 0.787(-06) & 0.846(-07) & 0.206(-08) \\
19.90&0.157(-05) & 0.687(-06) & 0.366(-07) & 0.321(-09) \\
14.65&0.213(-05) & 0.491(-06) & 0.826(-08) & 0.133(-10) \\
 9.41&0.283(-05) & 0.195(-06) & 0.337(-09) & 0.183(-13) \\
 \noalign{\smallskip}
  \hline
\end{tabular}
\label{Tab5}
\end{center}
\end{table}

As it follows from the radiative transfer equation the Rayleigh-Jeans brightness temperature for thermal (th) emission can be obtained from the next expression
\begin{eqnarray}
\delta T^{th}_{br,ul}&=&\frac{h\nu_{obs,ul}}{k_B}\left(\frac{1}{e^{h\nu_{ul}/k_BT_{ex}}-1} \right.\nonumber\\
&&\left.-\frac{1}{e^{h\nu_{ul}/k_BT_r}-1}\right)\left(1-e^{-\tau_{ul}}\right),
\end{eqnarray}
where $\nu_{obs,ul} = \nu_{ul}/(1+z)$ is the observed frequency of $\nu_{ul}$ line emission from halo placed at redshift z. Results of computations $\delta T^{th}_{br,ul}$ for halos with mass $M_h=5.3\cdot10^9$ M$_\odot$ are presented in Tab.~\ref{Tab6}, for other masses in Tab.~\ref{Tab8} in Appendix.

\begin{table}[h!]
\begin{center}
\caption{The thermal brightness temperatures $\delta T^{th}_{br}$ [K] and spectral flux density $\delta F_{ul}^{th}$ [$\rm{\mu Jy}$] in rotational lines of molecular ion HeH$^{+}$ 
for dark ages halos with mass $M=5.3\cdot 10^{9}$~$M_{\odot}$ formed at $z\simeq 10-30$. Powers of 10 are given in parentheses.}
\begin{tabular} {ccccc}
\hline
\hline
   \noalign{\smallskip}
$z$& &$T_{br}^{th}$[K]/$\delta F^{th}$[$\mu$Jy]& &  \\
 \noalign{\smallskip} 
&0-1&1-2&2-3\\
 \noalign{\smallskip} 
\hline
   \noalign{\smallskip} 
30.41&3.0(-15)/2.9(-11)&9.0(-16)/3.5(-11)&1.6(-16)/1.4(-11)\\
25.15&2.6(-15)/3.8(-11)&6.1(-16)/3.6(-11)&8.0(-17)/1.1(-11)\\     
19.90&2.0(-15)/4.9(-11)&3.3(-16)/3.2(-11)&2.6(-17)/5.6(-12)\\  
14.65&1.2(-15)/5.9(-11)&1.0(-16)/2.0(-11)&2.9(-18)/1.3(-12)\\
9.41&3.7(-16)/5.0(-11)&6.1(-18)/3.3(-12)&1.3(-20)/1.5(-14)\\  
 \noalign{\smallskip}
  \hline
\end{tabular}
\label{Tab6}
\end{center}
\end{table}

\begin{table}[h!]
\begin{center}
\caption{The resonant brightness temperatures $\delta T^{rs}_{br}$ [K] and spectral flux density $\delta F_{ul}^{rs}$ [$\rm{\mu Jy}$] in rotational lines of molecular ion HeH$^{+}$ for dark ages halos with mass $M=5.3\cdot 10^{9}$~$M_{\odot}$ formed at $z\simeq 10-30$. Powers of 10 are given in parentheses.}
\begin{tabular} {ccccc}
\hline
\hline
   \noalign{\smallskip}
$z$& &$T^{rs}_{br}$[K]/$\delta F^{rs}$[$\mu$Jy] & &  \\
 \noalign{\smallskip} 
&0-1&1-2&2-3\\
 \noalign{\smallskip} 
\hline
   \noalign{\smallskip}  
30.41&2.0(-7)/1.9(-3)&1.3(-7)/4.9(-3)&1.4(-8)/1.2(-3)\\
25.15&2.0(-7)/3.0(-3)&8.5(-8)/5.0(-3)&4.9(-9)/6.4(-4)\\     
19.90&1.9(-7)/4.8(-3)&4.4(-8)/4.4(-3)&9.4(-10)/2.1(-4)\\  
14.65&1.6(-7)/8.0(-3)&1.3(-8)/2.5(-3)&5.2(-11)/2.3(-5)\\
9.41&9.0(-8)/1.2(-2)&7.8(-10)/4.2(-4)&1.1(-13)/1.3(-7)\\  
 \noalign{\smallskip}
  \hline
\end{tabular}
\label{Tab7}
\end{center}
\end{table}
Since the halos in the Dark Ages possess the peculiar velocities $\vec{v}_{p}$ there are resonant scattering (rs) of CMB quanta that leads to the differential brightness temperature in rotational lines of HeH$^{+}$ ions \cite{Maoli1996,Persson2010}
\begin{equation}
 \delta T^{rs}_{br,ul}=\frac{h^2\nu_{ul}^2}{k_B^2T_r}\frac{e^{h\nu_{ul}/k_BT_r}}{(e^{h\nu_{ul}/k_BT_r}-1)^2}\left(1-e^{-\tau_{ul}}\right)\frac{|v_{p\shortparallel}|}{c},
\end{equation}
where $v_{p\shortparallel}$ is the projection  of the vector of peculiar velocity on the line of sight of Earth observer. 
A reliable estimation of peculiar velocity of halo with mass $M_h$ is the rms value  $V_{rms}\equiv \sqrt{<\vec{v}_p^2>}$, where
\begin{equation}
 <\vec{v}_p^2> = \frac{H^2(z)}{2\pi^2(1+z)^2}\int\limits_{0}^{\infty}P(k; z)W^2(kR)dk
 \label{v2}
\end{equation} 
where $P(k; z)$ is the power spectrum of density perturbations \cite{Eisenstein1998}, $W(x) \equiv 3(\sin x - x \cos x)/x^3$ is the top-hat sphere in Fourier space, and $R = (3M_h/4\overline{\rho}_m)^{1/3}$ is the comoving radius of the halo. Results of computations $\delta T^{th}_{rs,ul}$ for halos with mass $M_h=5.3\cdot10^9$ M$_\odot$ are presented in Tab.~\ref{Tab7}, for other masses in Tab.~\ref{Tab8} in Appendix.

We can also estimate the differential energy flux per unit frequency as follows \cite{Iliev2002,Novosyadlyj2019}
\begin{equation}
 \delta F_{ul} = 2\pi \left(\frac{\nu_{obs,ul}}{c}\right)^2k_B<\delta T_{br,ul}> \theta_h^2,
\end{equation}
where $\theta_h = r_h(M_h)/D_A(z)$ is the angular radius of halo with mass $M_h$, $D_A(z)$ is the angular diameter distance to the halo placed at redshift $z$. The halos analyzed here have angular radiuses $\sim$0.06"-1".
The results for thermal luminescence and resonant scattering for halos with mass $M_h=5.3\cdot10^9$ M$_\odot$ are presented in Tab.~\ref{Tab6} and Tab.~\ref{Tab7} correspondingly.

\begin{figure*}
\begin{center}
\animategraphics[autoplay,loop,width=0.99\linewidth]{1.0}{figure_H2}{0}{4}
\vskip-6.5cm
\caption{Evolution of opacity (left column) and thermal brightness temperature (right column) in the lines of transitions $J=1\rightarrow J=0$ (upper row), $J=2\rightarrow J=1$ (middle row) and $J=3\rightarrow J=2$ (bottom row) of helium hydride ion for halos with mass $M_h=5.3\cdot10^9$ M$_\odot$,   $M_h=6.6\cdot10^8$ M$_\odot$,  $M_h=8.3\cdot10^7$ M$_\odot$, $M_h=1.0\cdot10^7$ M$_\odot$, and $M_h=1.3\cdot10^6$ M$_\odot$ (animated). Each line corresponds to the halo with different initial amplitude of curvature perturbation: $C_k=3\cdot10^{-4},\,2.5\cdot10^{-4},\,2\cdot10^{-4},\,1.5\cdot10^{-4},\,1\cdot10^{-4},\,0$ (from top to bottom in the right hand side of each panel).  }
\end{center}
\label{fig1}
\end{figure*}

In Fig.~9 we present the evolution of opacity (left column) and thermal brightness temperature (right column) in the lines of rotational transitions $J=1\rightarrow J=0$ (upper row), $J=2\rightarrow J=1$ (middle row) and $J=3\rightarrow J=2$ (bottom row) for halos with mass $M_h=5.3\cdot10^9$ M$_\odot$, $M_h=6.6\cdot10^8$ M$_\odot$, $M_h=8.3\cdot10^7$ M$_\odot$, $M_h=1.0\cdot10^7$ M$_\odot$ and $M_h=1.3\cdot10^6$ M$_\odot$ (animated). Each line corresponds to the halo with different initial amplitude of curvature perturbation: $C_k=3\cdot10^{-4},\,2.5\cdot10^{-4},\,2\cdot10^{-4},\,1.5\cdot10^{-4},\,1\cdot10^{-4},\,0$ (from top to bottom in the right hand side of each panel and from bottom to top in the left hand side of each panel) which correspond to different times of halos virializations. As can be seen from the figure, the thermal brightness temperature increases due to increasing of the temperature of baryonic matter during compression of the proto-halo and decreases after the moment of halo virialization due to decreasing in the number of HeH$^{+}$ molecular ions. The similar peak-like time dependence is also expected for the resonance brightness temperature, but with an amplitude of approximately eight orders of magnitude greater (see Tab.~\ref{Tab8} of electronic version of this paper). We also note that an increase in the resonance brightness temperature in time caused by the increase in the amplitude of the peculiar velocity according to (\ref{v2}), and do not connect to an increase in the temperature of the baryon gas.

\section{Conclusions} \label{sec:cite}

In order to analyze the thermal luminescence and resonant scattering of CMB quanta in the rotational lines of molecule HeH$^+$ in dark ages halos
we have computed the cross-sections and rate coefficients of excitation/de-excitation of the lowest five rotational energy states of HeH$^+$ by
collisions with atomic hydrogen. We have shown also that in the conditions of Dark Ages the collision excitation/de-excitation of the lowest
rotational levels of HeH$^+$ by atoms of neutral hydrogen and electrons are competitive.

It was shown, due to the small number density of residual electrons in the Universe after recombination in redshifts range from 100 to 10, collisional excitation/de-excitation of the lowest rotational levels of ion-molecule HeH$^+$ by electrons and atomic hydrogen are comparable and both must be taken into account.
In the virialized halos, collisions with atomic hydrogen play a much more important role in the excitations and de-excitations of rotational levels of HeH$^{+}$ than collisions with electrons but both are subdominant in comparison with radiative activations/de-activations. Due to the peak time dependence of the thermal and resonance luminosities of the halo in HeH$^{+}$ lines, we can only observe those that have just formed at a given redshift, while the observations in lines of HD and H$_{2}$ \cite{Novosyadlyj2019} will give us an integral number of halos formed up to a given redshift. But extremely low differential brightness temperatures caused by thermal luminescence in rotational lines of HeH$^{+}$ caused only by electron and hydrogen collisional excitations do not leave a chance to detect them in the coming years. At the same time, the signal caused by resonant scattering of CMB quanta in the rotational lines of HeH$^{+}$ in dark ages halos may be detected by the next-generation telescopes.

Since, the rotational transitions in HeH$^{+}$ are more sensitive to the electromagnetic radiation than to collisions with electrons and hydrogen atoms the appearance in the early Universe of any other  electromagnetic radiation with another color temperature will radically increase the luminescence of these molecules and will make visible of dark ages halos against the cosmic microwave background, that will be studied in other paper.

\acknowledgments

This work was supported by the International Center of Future Science of Jilin University (P.R.China)
and the project of Ministry of Education and Science of Ukraine ``Formation and characteristics of elements of
the structure of the multicomponent Universe, gamma radiation of supernova remnants and observations of variable stars'' (state registration number 0119U001544).
%\newpage

\onecolumngrid
\section*{Appendix}
\appendix
The best-fit coefficients of analytical approximations for collisional de-excitation of four lowest rotational levels of HeH$^+$ molecule and the critical electron/hydrogen density for them are presented in the Tables \ref{Tab2} and \ref{Tab3} accordingly. The physical values and chemical composition of halos with mass $1.3\cdot10^6$, $1.0\cdot10^7$, $1.0\cdot10^7$, $8.3\cdot10^7$ and $6.6\cdot10^8$ M$_\odot$ and different initial amplitudes are presented in Table \ref{Tab4b}. The optical depths, brightness temperatures and spectral fluxes in the three lowest rotational lines of molecular ion HeH$^{+}$ from the dark ages halos of different masses and initial amplitudes are presented in \ref{Tab8}.

\begin{table}[h!]
\caption{Parameters $a_0$, $a_1$, $a_2$, $a_3$, $a_4$ and $a_5$ of the polynomial 
approximation (\ref{k_ul_approximation}) for the de-excitation rate coefficients 
of rotational transitions in molecular ion HeH$^{+}$ driven by collisions with atomic hydrogen.}
\label{Approx_coef}
\centering
\begin{tabular}{cccccccc}
\hline\hline
$j_u \to j_l$ & $\frac{\epsilon_u - \epsilon_l}{k_B}$,~[K] & $a_0$ & $a_1$ & $a_2$ & $a_3$ & $a_4$ & $a_5$\\
\hline
  1 -- 0 & 96  & 0.73387085  &  4.73875362  & -4.13010272 &  1.12333376 & -0.08995999 & -0.002472586  \\
  2 -- 0 & 289 & 0.70341481  &  1.65767339  & -1.71677310 &  0.54840014 & -0.06727353 &  0.002189463 \\
  3 -- 0 & 576 & 0.94178360  & -0.30602698  & -0.14440932 &  0.07307436 & -0.00731803 & -0.000220779 \\
  4 -- 0 & 956 & 0.69277704  & -0.89628132  &  1.20242960 & -0.77201220 &  0.20914845 & -0.020095280 \\
  2 -- 1 & 192 & 8.55444497  & -4.85218002  &  0.44909807 &  0.16516953 & -0.02689928 &  0 \\
  3 -- 1 & 480 & 2.49892140  &  2.69925750  & -3.33883170 &  1.03715820 & -0.10271530 &  0 \\
  4 -- 1 & 860 & 1.97539137  & -1.67054928  &  0.89376880 & -0.28221961 &  0.03374395 &  0 \\
  3 -- 2 & 287 & 5.84347772  & -2.27118822  & -0.18568545 &  0.17008179 & -0.01824961 &  0 \\
  4 -- 2 & 668 & 2.20317927  &  1.80004441  & -2.27352914 &  0.66729570 & -0.06129941 &  0 \\
  4 -- 3 & 380 & 5.259207601 & -3.33820790  &  0.77217303 & -0.08112202 &  0.00369328 &  0 \\
\hline 
\end{tabular}
\label{Tab2}
\end{table}

\begin{table}[h!]
\caption{Critical electron/hydrogen density (cm$^{-3}$) as a function of gass temperature for rotational levels $j=1,2,3,4$. Powers of 10 are given in parentheses.}
\label{critical_densities}
\centering
\begin{tabular}{cccccc}
\hline\hline
T, [K] & $j = 1$ & $j = 2$ & $j = 3$ & $j = 4$  \\
\hline
  10  &  9.37(4)/4.57(8) & 5.89(5)/2.40(9) & 2.28(6)/1.05(10)  & 6.94(6)/3.46(10) \\
  20  &  1.37(5)/5.16(8) & 9.04(5)/3.16(9) & 3.33(6)/1.29(10) & 9.44(6)/4.44(10) \\
  30  &  1.72(5)/5.79(8) & 1.16(6)/3.81(9) & 4.16(6)/1.48(10) & 1.14(7)/5.23(10) \\
  50  &  2.29(5)/6.89(8) & 1.59(6)/4.85(9) & 5.52(6)/1.78(10) & 1.45(7)/6.44(10) \\
  100 &  3.36(5)/9.30(8) & 2.44(6)/6.67(9) & 8.13(6)/2.31(10) & 2.04(7)/8.40(10) \\
  200 &  4.97(5)/1.40(9) & 3.73(6)/9.83(9) & 1.20(7)/3.09(10) & 2.90(7)/1.14(11) \\
  300 &  6.24(5)/1.88(9) & 4.78(6)/1.30(10)& 1.51(7)/3.78(10) & 3.59(7)/1.42(11) \\
  500 &  8.34(5)/2.81(9) & 6.51(6)/1.94(10)& 2.03(7)/5.07(10) & 4.73(7)/1.94(11) \\
  1000 & 1.24(6)/4.86(9) & 9.87(6)/3.38(10)& 3.03(7)/8.08(10) & 6.95(7)/3.10(11) \\
  2000 & 1.84(6)/8.31(9) & 1.49(7)/5.75(10)& 4.56(7)/1.40(11) & 1.04(8)/5.31(11) \\
\hline 
\end{tabular}
\label{Tab3}
\end{table}

\newpage

\begin{table*}[!ht]
\begin{center}
\caption{Physical values and chemical composition of halos of different mass and initial amplitudes\footnote{$M_h$ is the mass of halo, $k$ is the wave number of initial perturbation (seed of halo) from which the halo is formed, $C_k$ is the amplitude of initial curvature perturbation, $z_v$ is the redshift of virialization, $\rho_{m}$ is the matter density virialized halo, $T_K$ is kinetic temperature of baryonic gas, $n_{\rm HI}$ is the number density of neutral hydrogen atoms, $n_{\rm p},\,n_{\rm e}$ are the number densities of protons and electrons at $z=z_v/z=10$, $n_{\rm HeH^{+}}$ is the number densities of molecular ion HeH$^{+}$, $r_h$ is the radius of halo in comoving coordinates, $\theta_h$ is the angular radius of geometrically limited halo.}.}
\begin{tabular} {ccccccccccc}
\hline
\hline
   \noalign{\smallskip}
$M_h$&$k$&$C_k$&$z_v$&$\rho_{m}$&$T_K$&$n_{\rm HI}$&$n_{\rm p}\approx n_{\rm e}$&$n_{\rm HeH^{+}}$&$r_h$&$\theta_h$\\
 \noalign{\smallskip} 
[M$_{\odot}$]&[Mpc$^{-1}$]&[$10^{-4}$] & &[$10^{-24}$g/cm$^3$]&[K]&[cm$^{-3}$]&[10$^{-6}$cm$^{-3}$]&[10$^{-15}$cm$^{-3}$]&[kpc]&[arcsec]\\
 \noalign{\smallskip} 
\hline
$1.3(+06)$&80&3.0&50.33&66.6&834.0&5.00&355.3/3.6&8.473 &0.068&0.061\\
              &  &2.5&41.52&37.8&636.5&2.84&218.8/3.7&5.972 &0.082&0.062\\
              &  &2.0&32.65&18.7&446.8&1.41&117.7/3.7&3.835 &0.10 &0.064\\
              &  &1.5&24.40&8.07&286.1&0.61& 58.6/3.9&2.387 &0.14 &0.066\\
              &  &1.0&15.38&2.16&134.4&0.16& 17.1/4.2&1.014 &0.21&0.071\\ 
  \noalign{\smallskip}             
$1(+07)$&40&3.0 &45.72&50.1&730.3&3.76&287.2/3.6&7.336 &0.15&0.12\\
              & &2.5&37.85&28.8&556.0&2.16&179.7/3.7&5.198 &0.18&0.13\\
              & &2.0&29.92&14.6&392.9&1.09& 99.7/3.8&3.463 &0.23&0.13\\
              & &1.5&22.02&6.00&242.8&0.45& 45.7/3.9&2.021 &0.30&0.13\\
              & &1.0&14.00&1.66&115.2&0.13&14.1/4.3 &0.904 &0.47&0.14\\
    \noalign{\smallskip}            
 $8.3(+7)$&20&3.0&40.57&35.4&619.2&2.65&214.5/3.7&5.935 &0.34&0.25\\
              &  &2.5&33.55&20.3&468.4&1.52&133.5/3.7&4.246 &0.40&0.25\\
              &  &2.0&26.54&10.3&325.5&0.77& 74.0/3.8&2.824 &0.51&0.26\\
              &  &1.5&19.50&4.24&199.6&0.32& 33.8/4.0&1.645 &0.68&0.27\\
              &  &1.0&12.36&1.17& 94.0&0.09& 10.4/4.5&0.735 &1.05&0.30\\
    \noalign{\smallskip}           
$6.6(+08)$&10&3.0&35.54&24.0&508.9&1.80&156.3/3.7&4.771 &0.77&0.50\\ 
              &  &2.5&29.41&13.8&382.8&1.04& 97.4/3.8&3.428 &0.92&0.52\\
              &  &2.0&23.28&7.05&266.3&0.53& 54.2/3.9&2.287 &1.15&0.53\\
              &  &1.5&17.17&2.95&162.1&0.22& 25.1/4.2&1.359 &1.54&0.56\\
              &  &1.0&11.05&0.86& 78.2&0.06&  8.3/4.7&0.644 &2.32&0.61\\
 \hline 
\end{tabular}
\label{Tab4b}
\end{center}
\end{table*}

\begin{longtable}{ccccccccc}
\caption{The optical depths, brightness temperatures and spectral fluxes in the three lowest rotational lines of molecular ion HeH$^{+}$ in the dark ages halos of different masses $M_h$ virialized at different redshift $z_v$. Marking (th) means the thermal emission, marking (rs) means the resonant scattering. Powers of 10 are given in parentheses.} \\
\hline
\hline
$M_h$&$z_v$&$\nu_{obs}$&$\Delta\nu_{obs}$&$\tau_{\nu}$&$\delta T_{br}^{(th)}$&$\delta F^{(th)} $&$\delta T_{br}^{(rs)}$&$\delta F^{(rs)} $  \\
\noalign{\smallskip}
[M$_{\odot}$]& &[GHz]&[MHz]& &[K] &[$\mu$Jy]&[K] &[$\mu$Jy]  \\
\hline
\endfirsthead
\multicolumn{9}{c}%
{\tablename\ \thetable\ -- \textit{Continued from previous page}} \\
\hline
$M_h$&$z_v$&$\nu_{obs}$&$\Delta\nu_{obs}$&$\tau_{\nu}$&$\delta T_{br}^{(th)}$&$\delta F^{(th)} $&$\delta T_{br}^{(rs)}$&$\delta F^{(rs)} $  \\
\noalign{\smallskip}
[M$_{\odot}$]& &[GHz]&[MHz]& &[K] &[$\mu$Jy]&[K] &[$\mu$Jy]  \\
\hline
\endhead
\hline 
\multicolumn{9}{r}{\textit{Continued on next page}} \\
\endfoot
\hline
\endlastfoot
 \noalign{\smallskip} 
 \noalign{\smallskip} 
%\hline
  \noalign{\smallskip}
   1.3(+06)  & 50.33 & 39.2 & 0.36 & 0.426(-07) & 0.223(-15) & 0.288(-14) & 0.103(-07) & 0.133(-06) \\
             &       & 78.1 & 0.72 & 0.643(-07) & 0.105(-15) & 0.538(-14) & 0.139(-07) & 0.713(-06) \\
             &       & 117  & 1.07 & 0.288(-07) & 0.330(-16) & 0.378(-14) & 0.516(-08) & 0.591(-06) \\             
             & 41.52 & 47.3 & 0.38 & 0.555(-07) & 0.207(-15) & 0.403(-14) & 0.109(-07) & 0.214(-06) \\
             &       & 94.3 & 0.76 & 0.694(-07) & 0.843(-16) & 0.655(-14) & 0.116(-07) & 0.898(-06) \\
             &       & 141  & 1.13 & 0.226(-07) & 0.222(-16) & 0.384(-14) & 0.290(-08) & 0.502(-06) \\
             & 32.65 & 59.7 & 0.40 & 0.753(-07) & 0.178(-15) & 0.586(-14) & 0.114(-07) & 0.373(-06) \\
             &       & 119  & 0.80 & 0.711(-07) & 0.581(-16) & 0.760(-14) & 0.824(-08) & 0.108(-05) \\
             &       & 178  & 1.20 & 0.144(-07) & 0.116(-16) & 0.336(-14) & 0.112(-08) & 0.325(-06) \\             
             & 24.40 & 79.1 & 0.43 & 0.112(-06) & 0.147(-15) & 0.913(-14) & 0.119(-07) & 0.738(-06) \\
             &       & 158  & 0.85 & 0.695(-07) & 0.336(-16) & 0.829(-14) & 0.472(-08) & 0.116(-05) \\
             &       & 236  & 1.27 & 0.687(-08) & 0.417(-17) & 0.229(-14) & 0.240(-09) & 0.132(-06) \\             
             & 15.39 & 123  & 0.45 & 0.171(-06) & 0.711(-16) & 0.123(-13) & 0.944(-08) & 0.163(-05) \\
             &       & 245  & 0.91 & 0.440(-07) & 0.682(-17) & 0.469(-14) & 0.910(-09) & 0.625(-06) \\
             &       & 365  & 1.35 & 0.911(-09) & 0.234(-18) & 0.359(-15) & 0.491(-11) & 0.751(-08) \\
   \noalign{\smallskip}   
   1.0(+07)  & 45.72 & 43.0 & 0.37 & 0.100(-06) & 0.447(-15) & 0.284(-13) & 0.220(-07) & 0.139(-05) \\
             &       & 85.8 & 0.74 & 0.138(-06) & 0.197(-15) & 0.497(-13) & 0.263(-07) & 0.665(-05) \\
             &       & 128  & 1.10 & 0.533(-07) & 0.568(-16) & 0.320(-13) & 0.813(-08) & 0.458(-05) \\             
             & 37.85 & 51.7 & 0.39 & 0.130(-06) & 0.407(-15) & 0.392(-13) & 0.231(-07) & 0.223(-05) \\
             &       & 103  & 0.78 & 0.147(-06) & 0.153(-15) & 0.588(-13) & 0.214(-07) & 0.821(-05) \\
             &       & 154  & 1.16 & 0.404(-07) & 0.364(-16) & 0.311(-13) & 0.431(-08) & 0.369(-05) \\             
             & 29.92 & 65.0 & 0.41 & 0.177(-06) & 0.358(-15) & 0.569(-13) & 0.241(-07) & 0.383(-05) \\             
             &       & 130  & 0.82 & 0.149(-06) & 0.106(-15) & 0.670(-13) & 0.149(-07) & 0.941(-05) \\
             &       & 194  & 1.22 & 0.248(-07) & 0.186(-16) & 0.262(-13) & 0.155(-08) & 0.218(-05) \\             
             & 22.02 & 87.3 & 0.43 & 0.255(-06) & 0.268(-15) & 0.833(-13) & 0.237(-07) & 0.738(-05) \\
             &       & 174  & 0.87 & 0.133(-06) & 0.524(-16) & 0.648(-13) & 0.726(-08) & 0.898(-05) \\
             &       & 260  & 1.29 & 0.979(-08) & 0.524(-17) & 0.144(-13) & 0.243(-09) & 0.671(-06) \\             
             & 14.00 & 134  & 0.46 & 0.386(-06) & 0.125(-15) & 0.107(-12) & 0.182(-07) & 0.155(-04) \\
             &       & 267  & 0.92 & 0.800(-07) & 0.944(-17) & 0.320(-13) & 0.120(-08) & 0.406(-05) \\
             &       & 399  & 1.37 & 0.111(-08) & 0.219(-18) & 0.165(-14) & 0.358(-11) & 0.270(-07) \\
   \noalign{\smallskip}   
   8.3(+07)  & 40.57 & 48.4 & 0.38 & 0.237(-06) & 0.853(-15) & 0.280(-12) & 0.455(-07) & 0.149(-04) \\
             &       & 96.4 & 0.77 & 0.289(-06) & 0.341(-15) & 0.446(-12) & 0.465(-07) & 0.608(-04) \\
             &       & 144  & 1.14 & 0.903(-07) & 0.876(-16) & 0.255(-12) & 0.111(-07) & 0.322(-04) \\             
             & 33.55 & 58.2 & 0.40 & 0.306(-06) & 0.775(-15) & 0.384(-12) & 0.476(-07) & 0.236(-04) \\
             &       & 116  & 0.80 & 0.299(-06) & 0.260(-15) & 0.513(-12) & 0.362(-07) & 0.713(-04) \\
             &       & 173  & 1.20 & 0.644(-07) & 0.535(-16) & 0.235(-12) & 0.530(-08) & 0.233(-04) \\             
             & 26.54 & 73.0 & 0.42 & 0.415(-06) & 0.656(-15) & 0.542(-12) & 0.489(-07) & 0.404(-04) \\
             &       & 146  & 0.84 & 0.293(-06) & 0.168(-15) & 0.551(-12) & 0.236(-07) & 0.774(-04) \\
             &       & 217  & 1.25 & 0.363(-07) & 0.242(-16) & 0.177(-12) & 0.164(-08) & 0.120(-04) \\             
             & 19.50 & 98.0 & 0.44 & 0.584(-06) & 0.462(-15) & 0.752(-12) & 0.461(-07) & 0.750(-04) \\
             &       & 196  & 0.88 & 0.245(-06) & 0.736(-16) & 0.477(-12) & 0.998(-08) & 0.646(-04) \\
             &       & 292  & 1.32 & 0.122(-07) & 0.545(-17) & 0.786(-13) & 0.193(-09) & 0.278(-05) \\             
             & 12.36 & 150  & 0.47 & 0.849(-06) & 0.190(-15) & 0.855(-12) & 0.320(-07) & 0.144(-03) \\
             &       & 300  & 0.93 & 0.129(-06) & 0.981(-17) & 0.176(-12) & 0.121(-08) & 0.217(-04) \\
             &       & 448  & 1.39 & 0.997(-09) & 0.124(-18) & 0.497(-14) & 0.153(-11) & 0.610(-07) \\
   \noalign{\smallskip}   
   6.6(+08)  & 35.54 & 55.0 & 0.40 & 0.576(-06) & 0.164(-14) & 0.286(-11) & 0.957(-07) & 0.167(-03) \\
             &       & 110  & 0.79 & 0.604(-06) & 0.581(-15) & 0.404(-11) & 0.802(-07) & 0.558(-03) \\
             &       & 164  & 1.18 & 0.147(-06) & 0.128(-15) & 0.199(-11) & 0.137(-07) & 0.213(-03) \\             
             & 29.41 & 66.1 & 0.41 & 0.739(-06) & 0.145(-14) & 0.383(-11) & 0.986(-07) & 0.260(-03) \\
             &       & 132  & 0.82 & 0.606(-06) & 0.422(-15) & 0.443(-11) & 0.588(-07) & 0.618(-03) \\
             &       & 197  & 1.23 & 0.970(-07) & 0.719(-16) & 0.168(-11) & 0.579(-08) & 0.136(-03) \\             
             & 23.28 & 82.8 & 0.43 & 0.982(-06) & 0.118(-14) & 0.519(-11) & 0.983(-07) & 0.433(-03) \\
             &       & 165  & 0.86 & 0.564(-06) & 0.252(-15) & 0.441(-11) & 0.347(-07) & 0.608(-03) \\
             &       & 246  & 1.28 & 0.488(-07) & 0.284(-16) & 0.111(-11) & 0.147(-08) & 0.572(-04) \\
             & 17.17 & 111  & 0.45 & 0.137(-05) & 0.792(-15) & 0.684(-11) & 0.896(-07) & 0.773(-03) \\
             &       & 221  & 0.90 & 0.446(-06) & 0.987(-16) & 0.339(-11) & 0.128(-07) & 0.441(-03) \\
             &       & 329  & 1.34 & 0.141(-07) & 0.499(-17) & 0.382(-12) & 0.129(-09) & 0.989(-05) \\             
             & 11.05 & 167  & 0.47 & 0.193(-05) & 0.304(-15) & 0.705(-11) & 0.582(-07) & 0.135(-02) \\
             &       & 333  & 0.94 & 0.216(-06) & 0.105(-16) & 0.969(-12) & 0.126(-08) & 0.116(-03) \\
             &       & 497  & 1.40 & 0.938(-09) & 0.687(-19) & 0.141(-13) & 0.676(-12) & 0.139(-06) 
%\hline 
%\end{tabular}
\label{Tab8}
%\end{center}
\end{longtable}  
%\fi


\begin{thebibliography}{}
\bibitem{Aguado1992} A. Aguado and M. Paniagua, J. Chem. Phys. \textbf{96}, 1265 (1992).
\bibitem{Aquilanti2000} V. Aquilanti, G. Capecchi, S. Cavalli et al., Chem. Phys. Lett. \textbf{318}, 619 (2000).
\bibitem{Amaral2019} P.H.R. Amaral, L.G. Diniz, K.A. Jones, et al. Astrophys. J. \textbf{878}, 95 (2019).
\bibitem{Bovino2011} S. Bovino, M. Tacconi, F.A. Gianturco and D. Galli, Astron. Astrophys. \textbf{529}, A140 (2011).
\bibitem{Curik2017} R. Curik and C.H. Greene, J. of Chem. Phys. \textbf{147}, 054307 (2017).
\bibitem{Coppola2011} C.M. Coppola, L. Lodi and J. Tennyson, Mon. Not. R. Astron. Soc.  \textbf{415}, 487 (2011).
\bibitem{Dalgarno2005} A. Dalgarno, J. of Phys.: Conf. Ser. \textbf{4}, 10 (2005).
\bibitem{Dubrovich1977} V.K. Dubrovich,  Astron. Let. \textbf{3}, 66 (1977).
\bibitem{Dubrovich1997} V.K. Dubrovich, Astron. Astrophys. \textbf{324}, 27 (1997).
\bibitem{Dubrovich1995} V.K. Dubrovich and A.A. Lipovka, Astron. Astrophys. \textbf{296}, 307 (1995).
\bibitem{Dabrowski1978} I. Dabrowski and G. Herzberg, Ann. NY Acad. Sci. \textbf{38}, 14 (1978).
\bibitem{Engel2005} E.A. Engel, N. Doss, G.J. Harris and J. Tennyson, Mon. Not. R. Astron. Soc.  \textbf{357}, 471 (2005).
\bibitem{Eisenstein1998} D.J. Eisenstein and W. Hu, Astrophys. J. \textbf{496}, 605 (1998).
\bibitem{Fazio2012} D. De Fazio, M. de Castro-Vitores, A. Aguado, et al. J. Chem. Phys. \textbf{137}, 244306 (2012). 
\bibitem{Frommhold1978} L. Frommhold and H.M. Pickett, Chem. Phys. \textbf{28}, 441 (1978).
\bibitem{Galli1998} D. Galli and F. Palla, Astron. Astrophys. \textbf{335}, 403 (1998).
\bibitem{Galli2013} D. Galli and F. Palla, Ann. Rev. Astron. Astrophys. \textbf{51}, 163 (2013).
\bibitem{gloabel08} S.C.O. Glover and T. Abel, Mon. Not. R. Astron. Soc. \textbf{388}, 1627 (2008).
\bibitem{glosav08}  S.C.O. Glover, D.W. Savin, Mon. Not. R. Astron. Soc.  \textbf{393}, 911 (2009).
\bibitem{Gusten2019} R. Gusten, H. Wiesemeyer, D. Neufeld et al., Nature \textbf{568}, 357 (2019).
\bibitem{Hirata2006} C.M. Hirata and N. Padmanabhan, Mon. Not. R. Astron. Soc.  \textbf{372}, 1175 (2006).
\bibitem{izotov1984} Y.I. Izotov and I.G. Kolesnik, Soviet Astronomy \textbf{28}, 15 (1984).
\bibitem{Hunter1977} G. Hunter and M. Kuriyan, Proc. R. Soc. Lond. A \textbf{353}, 575 (1977).
\bibitem{Hogness1925} T.R. Hogness and E.G. Lunn, Phys. Rev. \textbf{26}, 44 (1925).
\bibitem{Molscat} M. Hutson and S. Green, MOLSCAT computer program, version 14, Distributed by Collaborative Computational Project No. 6 of the UK Engineering and Physical Sciences Research Council. (1994).
\bibitem{Hamilton2016} J.R. Hamilton, A. Faure and J. Tennyson, Mon. Not. R. Astron. Soc., \textbf{455}, 3281 (2016).
\bibitem{Iliev2002} I.T. Iliev, E. Scannapieco, H. Martel and P.R. Shapiro, Mon. Not. R. Astron. Soc.  \textbf{341}, 81 (2003).
\bibitem{Joseph1987} T. Joseph and N. Sathyamurthy, J. of Chem. Phys. \textbf{86}, 704 (1987).
\bibitem{Jura1971} M. Jura and A. Dalgarno, Astron. Astrophys. \textbf{11}, 243 (1971).
\bibitem{Jurek1995} M. Jurek, V. Spirko and W.P. Kraemer, Chem. Phys. \textbf{193}, 287 (1995).
\bibitem{Khamesian2018} M. Khamesian, M. Ayouz, J. Singh and V. Kokoouline, Atoms \textbf{6(3)}, 49 (2018).
\bibitem{Kimura1993} M. Kimura, N.F. Lane, A. Dalgarno and R.G. Dixson, Astrophys. J. \textbf{405}, 801 (1993).
\bibitem{Kraemer1995} W.P. Kraemer, V. Spirko and M. Jurek, Chem. Phys. Lett. \textbf{236}, 177 (1995).
\bibitem{Lepp2002} S. Lepp, P.C. Stancil and A. Dalgarno, J. Phys. B. \textbf{35}, R57 (2002).
\bibitem{Lesiuk2014} M. Lesiuk and R. Moszynski, Phys. Rev. E \textbf{90}, 063318 (2014).
\bibitem{Liu1997} X.W. Liu, M.J. Barlow, A. Dalgarno, J. Tennyson et al.,  Mon. Not. R. Astron. Soc.  \textbf{290}, L71 (1997).
\bibitem{McLaughlin1979} D.R. McLaughlin and D.L. Thompson, J. of Chem. Phys. \textbf{70}, 2748 (1979).
\bibitem{Michels1966} H.H. Michels, J. Chem. Phys. \textbf{44}, 3834 (1966).
\bibitem{Ming2015} M. Li and B. Gao, Phys. Rev. A \textbf{91}, 032702 (2015).
\bibitem{Maoli1996} R. Maoli, V. Ferrucci, F. Melchiorri F. and D. Tosti, Astrophys. J. \textbf{457}, 1 (1996).
\bibitem{Novosyadlyj2006} B. Novosyadlyj,  Mon. Not. R. Astron. Soc.  \textbf{370}, 1771 (2006).
\bibitem{Novosyadlyj2016} B. Novosyadlyj, M. Tsizh and Yu. Kulinich, Gen. Relativ. Grav., \textbf{48}, 30 (2016).
\bibitem{Novosyadlyj2018} B. Novosyadlyj, V. Shulga, W. Han, Yu. Kulinich and  M. Tsizh, Astrophys. J. \textbf{865}, 38 (2018).
\bibitem{Novosyadlyj2017} B. Novosyadlyj, O. Sergijenko and V. Shulga, Kin. and Phys. of Cel. Bod. \textbf{33}, 255 (2017).
\bibitem{Novosyadlyj2019} B. Novosyadlyj, V. Shulga, Yu. Kulinich and W. Han, Astrophys. J. \textbf{888}, 27 (2020).
\bibitem{Persson2010} C.M. Persson, R. Maoli, P. Encrenaz et al.,  Astron. Astrophys. \textbf{515}, A72 (2010).
\bibitem{pfe03} D. Pfenniger and D. Puy, Astron. Astrophys. \textbf{398}, 447 (2003).
\bibitem{Puy1993} D. Puy, G. Alecian, J. Le Bourlot, J. Leorat and G. Pineau Des Forets, Astron. Astrophys. \textbf{267}, 337 (1993).
\bibitem{Puy1996} D. Puy, and M. Signore, Astron. Astrophys. \textbf{305} 371 (1996).
\bibitem{Puy1999} D. Puy and M. Signore, New Astron. Rev. \textbf{43}, 223 (1999).
\bibitem{Puy2007} D. Puy and M. Signore M., New Astron. Rev. \textbf{51}, 411 (2007).
\bibitem{Palmieri2000} P. Palmieri, C. Puzzarini, V. Aquilanti et al., Mol. Phys. \textbf{98}, 1835 (2000).
\bibitem{Rabadan1998} I. Rabadan, B.K. Sarpal and J. Tennyson, Mon. Not. R. Astron. Soc.  \textbf{299}, 171 (1998).
\bibitem{Roberge1982} W. Roberge and A. Dalgarno, Astrophys. J. \textbf{255}, 489 (1982).
\bibitem{Ramachandran2009} C.N. Ramachandran, D. De Fazio, S. Cavalli, F. Tarantelli and V. Aquilanti, Chemical Physics Letters \textbf{469}, 26 (2009).
\bibitem{sta98} P.C. Stancil, S. Lepp, A. Dalgarno, Astrophys. J. \textbf{509}, 1 (1998).
\bibitem{vonlanthen2009} P. Vonlanthen, T. Rauscher, C. Winteler et al., Astron. Astrophys. \textbf{503} 47(2009).
\bibitem{Xu2008} W. Xu, X. Liu, S. Luan et al., Chem. Phys. Lett. \textbf{464}, 92 (2008).
\bibitem{Zinchenko2011} I. Zinchenko, V. Dubrovich, C. Henkel, Mon. Not. R. Astron. Soc. \textbf{415}, L78 (2011).
\bibitem{Zygelman1998} B. Zygelman, P.C. Stancil and A. Dalgarno, Astrophys. J. \textbf{508}, 151 (1998).
\end{thebibliography}
\end{document}